\documentclass{article}
\usepackage[utf8]{inputenc}
\usepackage{amsmath,amsthm,amssymb,amsfonts,mathtools,bm,bbm,array,float,mathdots,dsfont,mathrsfs}
\usepackage{caption,subcaption,multirow,longtable,lscape,wrapfig}
\usepackage{hyperref}
\hypersetup{allcolors=black}
\usepackage[backend=bibtex,style=ieee]{biblatex} 
\providecommand{\keywords}[1]
{
  \small	
  \textbf{Keywords:} #1
}

\begin{document}

\title{Machine Learning in Physics \& Geometry}

\author{Yang-Hui He\textsuperscript{1,2,3,4}\footnote{hey@maths.ox.ac.uk}, Elli Heyes\textsuperscript{2,1}\footnote{elli.heyes@city.ac.uk}, Edward Hirst\textsuperscript{2,1}\footnote{edward.hirst@city.ac.uk}}

\date{\textsuperscript{1}\textit{\small London Institute for Mathematical Sciences, Royal Institution, \\ London W1S 4BS, UK} \\
\textsuperscript{2}\textit{\small Department of Mathematics, City, University of London, \\ EC1V 0HB, UK} \\
\textsuperscript{3}\textit{\small Merton College, University of Oxford, OX1 4JD, UK} \\
\textsuperscript{4}\textit{\small School of Physics, NanKai University, Tianjin, 300071, P.R. China}
}

\maketitle
\numberwithin{equation}{subsection}
\numberwithin{figure}{subsection}
\numberwithin{table}{subsection}

\begin{abstract}
We survey some recent applications of machine learning to problems in geometry and theoretical physics. 
Pure mathematical data has been compiled over the last few decades by the community and experiments in supervised, semi-supervised and unsupervised machine learning have found surprising success. We thus advocate the programme of machine learning mathematical structures, and formulating conjectures via pattern recognition, in other words using artificial intelligence to help one do mathematics.

This is an invited chapter contribution to Elsevier's {\it Handbook of Statistics, Volume 49: Artificial Intelligence} edited by S.~G.~Krantz, A.~S.~R.~Srinivasa Rao, and C.~R.~Rao.
\end{abstract}


\tableofcontents


\section{Introduction \& Summary} 

The fields of mathematics and physics are intimately intertwined and have been so since the dawn of philosophy. The ancient Greeks studied the conic sections, curves obtained by slicing the cone by a plane. By Descartes, these were realized to be the geometry of quadratic polynomials in two variables. Finally, Newton combined his Second Law with the inverse square Law of Universal Gravitation to show that the solution to the motion of planets were precisely these conics. Gauss, Bolyai and Lobachevsky mused with an intrinsic, non-Euclidean geometry, and Riemann was able to formulate it into the foundations of differential geometry. Then, Einstein was able to marry this to his General Relativity and demonstrated that gravity was the curvature of spacetime.

Such examples abounded throughout the history of fundamental physics, where there is a perpetual dance between the nature of reality and the structures of mathematics. One could say that modern physics, from relativity to quantum field theory, has been following a ``Geometrization Programme''. Any unified theory must follow this path. String theory, still the best contender for this ``theory of everything'', is favoured despite the lack of experimental evidence thus far, precisely because it is a brain-child of this geometrization programme.

Traditionally, computations in mathematics and physics were done solely by hand. When computers first came about mathematicians and physicists were eager to use them to speed up calculations. Think back to the days when the likes of Euler had to find the digits of $\pi$ by hand! The development of computers has been tremendous and one can now carry out calculations on a personal laptop that would take years to complete by hand in just a matter of seconds. This clearly was a huge advantage for contemporary mathematicians and physicists. Imagine how much more Euler would have achieved if he had a symbolic package like Mathematica \cite{Mathematica} to help with the mundane tasks of obtaining integrals and series.

And this brings us to the heart of the matter. We often forget, when reading the final and polished journal papers, that theoretical research - even in pure mathematics - begins with {\it experimentation}. Gauss would plot the number $N(x)$ of primes less than a given real number $x$ and notice that it is approximated by $x / \log(x)$, long before complex analysis was invented to offer a proof. Einstein would add a $\Lambda$ constant to his field equations just to toy with a cosmological model long before the theories of inflation or evolution. This is what mathematicians {\it do}; we experiment and {\it then} we formalize. In the words of the great Vladimir Arnol'd, ``mathematics is a branch of physics where experiments are cheap.'' Henceforth, we will use the term ``mathematician'' to include pure and applied mathematicians, as well as theoretical physicists.

\subsection{Mathematical Data as Pure Data}

Here, then, lies the paradigmatic change. Whilst since at least the mid-twentieth century computers have aided mathematicians, only the last decade or so - considered the beginnings of the ``Age of Data'' - has brought in a plethora of ``mathematical data''. All of a sudden, any mathematician has free and instant access to Gigabytes of data from across the disciplines, and, importantly, the ability to experiment with this data on their laptops.
There is, by definition, something quite unique about mathematical data: distinct from data coming from the ``real world'', it is largely {\it noiseless}. Simulations of numerical solutions aside, pure mathematical data such as the parity of number of prime factors of the first one million integers, or the list of inequivalent graphs with a given property up to 100 vertices, are {\it exact}. Suddenly, we have an abundance of this sort of data at our fingertips and finding patterns in them, like what Gauss did with $N(x)$, leads to conjectures and mathematical discovery.

Treating large volumes of data inevitably brings us to the topic of artificial intelligence. In the past few decades, the field of machine learning (ML) (a subfield of artificial intelligence (AI)) has suddenly taken off and has become an integral part of scientific research. In contrast to traditional programming, where one tells the computer the exact steps to solve the problem, machine learning algorithms are able to solve problems without this instruction. Of course, AI and ML techniques have long been an indispensable tool in the experimental sciences; in fundamental physics, for instance, the Higgs particle could not have been found by CERN without deep-learning of the detector data. It might seem counter-intuitive that a methodology grounded upon the statistical inference of data should be of any use to the rigorous world of proofs in mathematics and derivations in theoretical
physics.

Nevertheless, from the introduction in 2017 of ML into the mathematics of algebraic geometry and the physics of superstring theory \cite{He:2017aed,Carifio:2017bov,Krefl:2017yox,Ruehle:2017mzq} (see also the usage of genetic algorithms in searching for the correct vacuum \cite{Abel:2014xta}), to the recent advances of Google's DeepMind collaboration with mathematicians in studying knot invariants in 2022 \cite{Davies2021}, as well as work to uncover physical laws directly \cite{AIFeyman}, there has been an explosion of activity in the past few years.

\subsection{The Inevitability of AI in Geometry \& Physics}

In this chapter we shall focus mainly on string theory as an example where ML is helping push the boundaries of physics and geometry and shall present some specific examples. String theory is currently the most promising candidate for a theory of everything (a theory that combines quantum mechanics and general relativity); it is a brain-child of the geometrization programme which modern physics has followed.

The main premise is that everything in the universe is built from a fundamental string object, whose motion gives rise to different particle properties. A requirement for string theory to be self-consistent is that there exists an extra six dimensions of space, on top of the three dimensions of space and one dimension of time that we are familiar with. The geometry of the extra dimensions together with the associated metric determines the physics of the resulting universe, such as the masses and charges of particles, the coupling between forces, etc. Therefore, the mathematical tools from geometry used to construct these spaces are crucial for string theorists.

The biggest theoretical challenge to string theory is that there is a large number of possible geometries that could describe the extra six dimensions and we have no method for choosing one that resembles our universe. This is called the vacuum degeneracy problem.
We refer to the collection of possible geometries as the ``string landscape''. The size of the landscape is certainly too big to study each by hand and even too big to exhaustively search using a computer. To give an idea, a small corner of the string landscape, consisting of so-called flux compatifications for Calabi-Yau manifolds, there are some $10^{500}$ possibilities \cite{Kachru:2003aw}. This is where ML can help. The above-mentioned references brought machine-learning and AI techniques to the 
search of the string landscape in order to identify regions that give rise to sensible physics.

As well as choosing a geometry from the landscape, we must also find the metric associated to the geometry, which is an equation which defines the notion of distance on the geometry. There is currently no method for the construction of an explicit metric on a Calabi-Yau. This is another area where ML has proved useful. There have been attempts to approximate an appropriate (Ricci flat) metric on a Calabi-Yau starting with direct numerical methods through Donaldson's algorithm \cite{Donaldson1,Donaldson2,Donaldson3}, as well as with reinforcement learning techniques \cite{Douglas:2006rr,Braun:2007sn,Ashmore:2019wzb}, where suitable packages are now available \cite{Larfors:2021pbb,gerdes2022cyjax}.

In this work we review a selection of common ML techniques from both supervised and unsupervised learning, and present some examples of their applications in physics and geometry. The examples focus on four areas of string theory and geometry, namely: Calabi-Yau manifolds from polytopes, amoebae, quivers, and brane webs. We begin in Section \ref{sec:background} by introducing the relevant physical and mathematical background. Section \ref{sec:supervised} focuses on supervised learning, in particular neural networks (NN) §\ref{sec:nn} and support vector machines (SVM) §\ref{sec:svm}. In Section \ref{sec:unsupervised} we look at unsupervised learning, in particular principal component analysis (PCA) §\ref{sec:pca}, t-distributed stochastic neighbourhood embedding (t-SNE) §\ref{sec:tSNE}, $k$-means clustering §\ref{sec:kmeans}, and topological data analysis (TDA) §\ref{sec:tda}. Finally leaving some concluding remarks in  Section \ref{sec:conc}.

\section{Background Physics and Mathematics}\label{sec:background} 

As technology developed throughout the twentieth century, the equipment used for experimentation became unimaginably more advanced. The scale of experiments hence greatly expanded, in both directions, as mankind could now probe both the deepest parts of space and the smallest scales of particles. However, comfortable with a theory of atoms, these experiments at shorter distance scales (and hence higher energy densities) instead of confirming the atomic theory saw an explosion of new observed particles, as well as new seemingly illogical effects.

It was clear that classical theory was no longer appropriate for describing these unusual effects occurring at small distance scales, and so from the ashes of atomic theory was born a new style of theory: quantum physics. At these small distance scales the continuous nature of the particle properties was lost, as energy levels became discrete and hence `quantised'; and over time as the theory of quantum physics was developed more of this unusual behaviour could be described. Despite this success, there was still desire to incorporate the ideas which so succinctly explained the physics of long distances into quantum theory, in particular Einstein's relativity.

Relativity, as a theory, was the result of an exceptionally insightful paradigm shift in how space and time were viewed.
Instead of a three-dimensional space ticking along through time, the playground on which our physical theory lives was re-imagined as a manifestly four-dimensional spacetime. Depending on our relative speeds, we saw the universe from a different angle (a different `inertial frame') and this could lead to discrepancies in experiment -- later confirmed through observation. Relativity's frames were hence based on speeds, with a maximum speed of light which could not be exceeded in any frame. As special relativity was developed into general relativity to incorporate accelerating objects and massive bodies, the foundations of gravity in geometry were cemented.
Inertial mass dictating acceleration was now seen as equivalent to gravitational mass, and introduction of mass warped the spacetime from a flat Minkowski manifold to a curved Lorentzian manifold.

All physical theories rely heavily on the concept of symmetry, described through the mathematical language of groups.
Knowing a theory's symmetry can greatly simplify calculations, and provides the means for a sensible development of the correct intuition. The symmetry of special relativity is based on the Poincar{\'e} group of spacetime translation and rotation. 
The rotation part is the Lorentz group, and knowing how the theory's fundamental objects transform under this group is essential for computation in the theory. For our four-dimensional spacetime this group is $SO(1,3)$, but it is turns out to be more natural to consider the double cover $Spin(1,3)$, where the theory's particles transform as representations of this group. These Lorentz representations end up being classified according to a particle property known as spin, which can take integer or half-integer values, categorising the particle as a boson or fermion respectively. The most common representations have lower spin values and hence their own special names: spin $0$ particles are scalars, spin $\frac{1}{2}$ particles are spinors, and spin $1$ particles are vectors. 

To incorporate the central ideas of special relativity into quantum theory the fundamental particle objects are changed to fields in particular representations of the Lorentz group, so birthing quantum field theory. Beyond the Poincar{\'e} symmetry somewhat associated to gravity, there are other symmetry groups related to the other three fundamental forces. The current best theory for our universe is the `standard model' in flat spacetime with symmetry group $SU(3) \times SU(2) \times U(1)$ to describe the strong nuclear force, the weak nuclear force, and electromagnetism respectively. 
The standard model is truly a \textit{gauge} theory, and in the gauging process the symmetries described by the given groups are generalised from global to local, and require the introduction of gauge vector bosons which moderate the respective interactions between the theory's matter fields. Ultimately, physicists seek a \textit{theory of everything}, which unalike the standard model, works at all distance scales describing all forces. The elusive piece for the standard model is gravity, as although the techniques of quantum field theory work well with special relativity, they fail with general relativity as gravity cannot be quantised.

The most famous candidate for a theory of everything is \textit{string theory}. String theory was developed out of another paradigm shifting idea, answering the question: why do the fundamental objects of our universe have to be 0-dimensional particles? Relaxing this assumption that the building blocks of the universe are particles (these 0-dimensional dots), a range of other fundamental objects can be included, such as 1-dimensional strings, 2-dimensional membranes, etc. As one may guess from the name, the most important objects in these theories at the low-energies of our reality turn out to be strings. String theory is importantly self-consistent to all energies, however as mentioned previously does require the introduction of extra spatial dimensions into spacetime.
In the limit of lower energies different string theories with different spacetime geometries reduce to different quantum field theories, and the task of quantising gravity is reaffirmed concretely within the realm of geometry.

With the aim of application of ML techniques to theoretical physics the first challenge is the identification and construction of relevant datasets. Each dataset consists of a set of mathematical objects of a particular type, which are represented in an appropriate way for computational processing. 
The type of mathematical object used (as well as the representation choice) may change the perspective with which the physical theory is viewed. Often is the case that these dual viewpoints offer non-trivial insights unavailable from the original representation.
This duality concept leads to a large multitude of objects which provide unique viewpoints on the same theory, a selection of which are chosen for data generation, analysis, and machine learning for the research work reviewed in this chapter. Further reviews of relevant work at the intersection of machine learning, physics and geometry can be found at \cite{He:2018jtw,Ruehle:2020jrk,He:2021oav,Bao:2020sqg,Bao:2022rup}.

\subsection{Polytopes}\label{sec:polytopes} 

The first mathematical dataset which we shall consider is that of polytopes. A polytope is a geometrical object that is bounded by a finite set of faces. Polytopes can exist in any number of dimensions, polytopes in 2 dimensions are also called polygons and polytopes in 3 dimensions polytopes are also called polyhedra. Simple examples of polytopes include the filled in triangle in 2 dimensions and the filled in cube in 3 dimensions.

Polytopes are of particular interest to string theorists because in 1993 it was shown \cite{batyrev1993} that one can obtain a Calabi-Yau (CY) manifold from a particular type of polytope, called a reflexive polytope, and CY manifolds play an important role in string theory as candidate geometries for the compactification space \cite{Candelas:1985en}. Roughly speaking, from a polytope $\Delta$ one can construct a geometric space (toric variety), and it happens that if $\Delta$ is reflexive then certain subspaces (hypersurfaces) of this space describe CY manifolds. 

\begin{figure}[!h]
    \centering
    \includegraphics[width=0.4\textwidth]{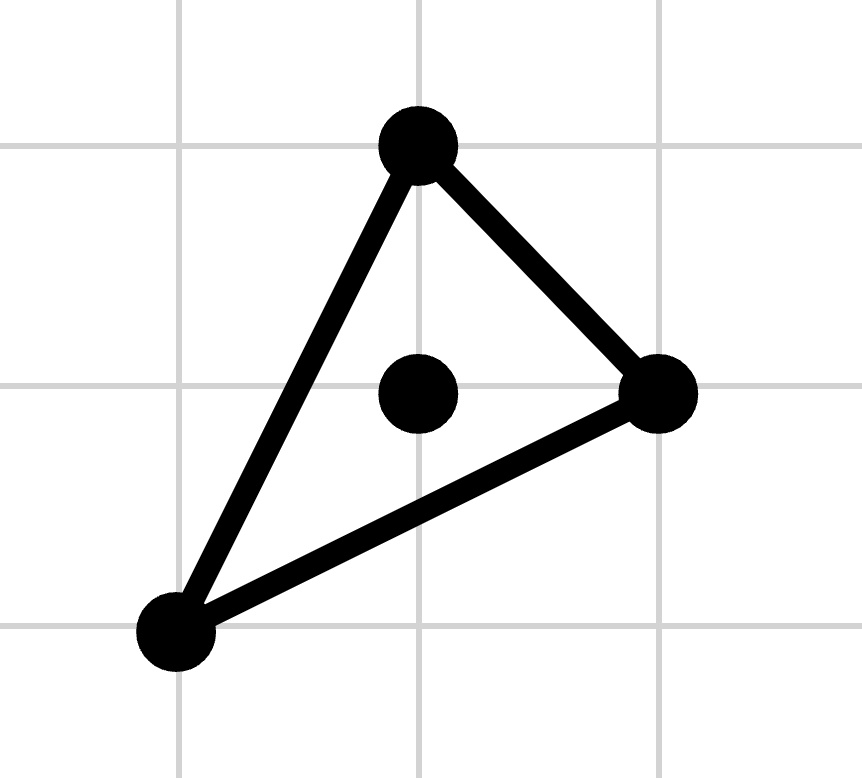}
    \caption{An example 2-dimensional reflexive polytope, with vertices $\{(1,0),(0,1),(-1,-1)\}$.}
    \label{polytope_example}
\end{figure}

Let's now define polytopes and the notion of reflexivity in more detail. Let $M$ and $N$ be a dual pair of $n$-dimensional integer lattices, meaning that lattice points of $M$ are described by a set of $n$ integers $X=(x_{1},...,x_{n})$, and lattice points of $N$ describe a linear map $f:M\rightarrow\mathbb{Z}$. There exists a natural pairing between the two lattices $\langle \cdot,\cdot \rangle: N \times M \rightarrow \mathbb{Z}$, where $\langle f,X \rangle = f(X)$. We define a polytope $\Delta$ in $M$ as the convex hull of a finite number of points. Such a polytope is called a lattice polytope if all its vertices lie on lattice points. The dual polytope $\Delta^{*}$ is defined as the set of all points $f$ in $N$ such that $\langle f,X \rangle = f(X) \geq -1$ for all points $X$ in $M$. A polytope is said to satisfy the interior point (IP) property when it contains only one interior point at the origin. Let $\Delta$ be a lattice polytope satisfying the IP property, then $\Delta$ is called reflexive if its dual $\Delta^{*}$ is also a lattice polytope satisfying the IP property. An example of a reflexive polytope in 2 dimensions is given in Figure \ref{polytope_example}.

Motivated by the discovery that CY manifolds can be built from reflexive polytopes, in 1997 Kreuzer and Skarke devised an algorithm to classify reflexive polytopes of any dimension \cite{kreuzer1997}. This algorithm consists of two steps: Firstly construct a set $S$ of ``maximal'' polytopes, such that any reflexive polytope is a subpolytope of a polytope in $S$. These maximal polytopes are defined by a weight system or a combination of weight systems. Secondly, given $S$, construct all subpolyhedra of all polyhedra in $S$ and check for reflexivity. The complete classification of 4319 3-dimensional reflexive polytopes by this algorithm was presented in \cite{kreuzer1998}. The 184,026 weight systems giving rise to 4-dimensional reflexive polytopes was presented in \cite{skarke1996} and the classification of 473,800,776 4-dimensional reflexive polytopes from these weight systems was given in \cite{Kreuzer:2000xy}.\footnote{Data of refleixve polytopes obtained by Kreuzer and Skarke's algorithm can be found on the website \cite{KSweb}.}. 

Included in the dataset of all 4-dimensional reflexive polytopes is the subset of polytopes which give rise to the weighted projective space $W\mathbb{P}^{4}$. There are 7555 such polytopes characterised by a set of weights $w_{i=0,...,4}$, first computed in \cite{Candelas:1989hd}. This database was analysed with ML techniques in \cite{Berman:2021mcw} and it was discovered that the topological data of the resulting CYs depend almost linearly on the weights. Furthermore, in \cite{Bao:2021ofk} it was shown that supervised ML techniques are able to predict with high accuracy standard properties of more general polytopes such as the volume, dual volume, reflexivity, etc. Later on we shall give explicit example from these two papers.

\subsection{Amoebae}\label{sec_amb} 

Amoebae, introduced by Gelfand, Kapranov and Zelevinsky in 1994 \cite{zelevinsky1994}, are regions in $\mathbb{R}^{n}$ with several holes and straight narrowing tentacles reaching to infinity, constructed from polynomials in $n$ complex variables. 

Considering some $n$-dimensional lattice polytope in $\mathbb{Z}^n$, one can construct a complex polynomial in $\mathbb{C}^n$, called the Newton polynomial. Specifically, each polytope point is associated to a monomial term in the polynomial equation, where each of the $n$ $\mathbb{C}^n$ coordinates in the respective monomial is raised to the power of the respective entry in the considered point's lattice coordinates. The amoeba is then generated through a log projection of the polynomial, using the polar representation of the complex space coordinates, $z_j = e^{(s_j+i\theta_j)}$, such that $Log: \ z_j \longmapsto s_j = log|z_j|$, overall mapping the polynomial from $\mathbb{C}^n$ to $\mathbb{R}^n$. Conversely, the algae map may be considered onto the $\theta_j$ coordinates \cite{Feng:2005gw}. 

\begin{figure}[!h]
    \centering
    \includegraphics[width=0.4\textwidth]{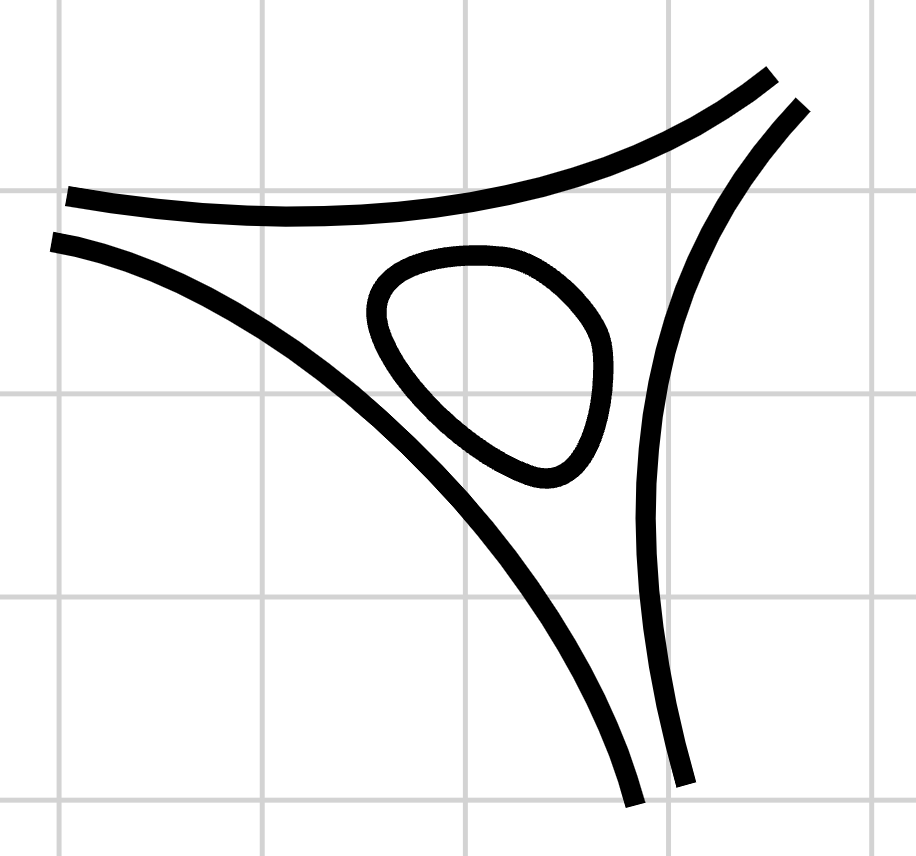} 
    \caption{An example amoeba constructed as the amoeba projection of the Newton polynomial for the lattice polytope with vertices $\{(1,0),(0,1),(-1,-1)\}$; varying the polynomial coefficients can change to hole's size/existence.}
    \label{amoeba_example}
\end{figure}

As an example, let's consider the 2-dimensional lattice polytope with coordinates $\{(1,0),(0,1),(-1,-1)\}$, as shown in Figure \ref{polytope_example}. To construct the Newton polynomial, each of these three coordinates contributes a monomial term defined using the equivalent $\mathbb{C}^2$ space parameterised by $(z_1,z_2)$.
This gives polynomial equation:
$0 = c_\alpha z_1^1z_2^0 + c_\beta z_1^0z_2^1 + c_\gamma z_1^{-1}z_2^{-1} = c_\alpha z_1 + c_\beta z_2 + \frac{c_\gamma}{z_1z_2}$, where each $c_{i}$ is some complex coefficient. After log projection we get the resulting amoeba, drawn in Figure \ref{amoeba_example}. 

This amoeba projection provides a visualisation tool for examining the modulus properties of the complex polynomial. Importantly, how the map projects the polynomial's topology will change dependent on the choice of coefficients for the polynomial. The complex polynomial will have a particular genus, and each of the surface's holes may or may not be projected under the amoeba map to the respective amoeba. This leads to certain bounds in the coefficient space, where different coefficient values will lead to an amoeba with different genus values, but determining these bounds is computationally demanding. It is well understood that ML techniques such as CNNs are very good image processing. This therefore raises the question, could ML determine the bounds in coefficient space from images of amoeba? This is precisely the topic of \cite{Bao:2021olg,Chen:2022jwd}. We will discuss the results of this paper later on in this review.

\subsection{Quivers}\label{sec:quivers} 

\begin{figure}[!h]
    \centering
    \includegraphics[width=0.4\textwidth]{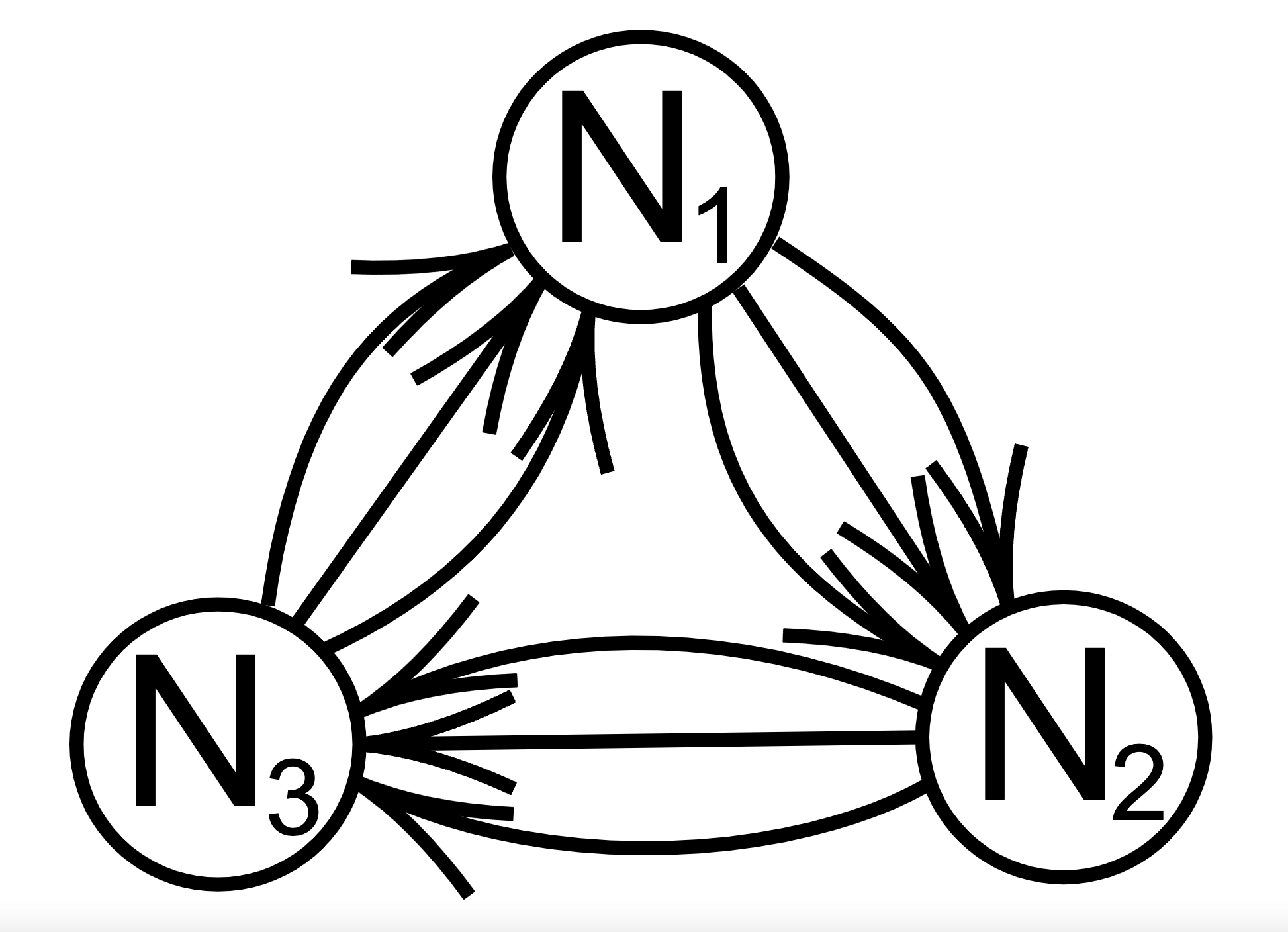}
    \caption{The trinian quiver, representing an $\mathcal{N}=1$ supersymmetric gauge theory with three $U(N_i)$ gauge groups, each with an implicit vector multiplet from gauging, and three chiral multiplets connecting each pair of gauge groups. For a specific superpotential, and where all of the $N_i=1$, the quiver theory's moduli space is the toric variety constructed from the lattice polytope with vertices $\{(1,0),(0,1),(-1,-1)\}$.}
    \label{quiver_example}
\end{figure}

In 1972, Gabriel introduced the concept of a quiver \cite{gabriel1971}, which is simply a multi-digraph consisting of a set of nodes connected by directed edges. This sparked a plethora of research in many branches of mathematics. Physicists also discovered that these objects can be used to study gauge theories \cite{Nakajima:1994nid,Douglas:1996sw}. They found that the gauge fields and matter content of some quantum field theories can be neatly encoded in a quiver, where each node is a respective $U(n_{i})$ symmetry group of the full theory with an implicit vector multiplet arising from the gauging of the group, whilst each edge is a chiral multiplet transforming under the symmetries of the two nodes it is connected to. An example quiver is shown in Figure \ref{quiver_example}.

Infinitely many quivers can be drawn, each equivalent to some supersymmetric gauge theory. However which of these can describe the universe we live in? To probe this question we must examine the low-energy behaviour of the theory, at scales reachable by our experimental technology. During the process of reducing the energy of a theory there is a limiting point, called an infra-red fixed point, below which the physics is fixed. One can categorise the infinite space of potential quivers, into groups based on having the same infra-red fixed point. Given one quiver gauge theory it is easy to generate another infra-red equivalent theory through a process known as quiver mutation.

\begin{figure}[!h]
    \centering
    \includegraphics[width=0.6\textwidth]{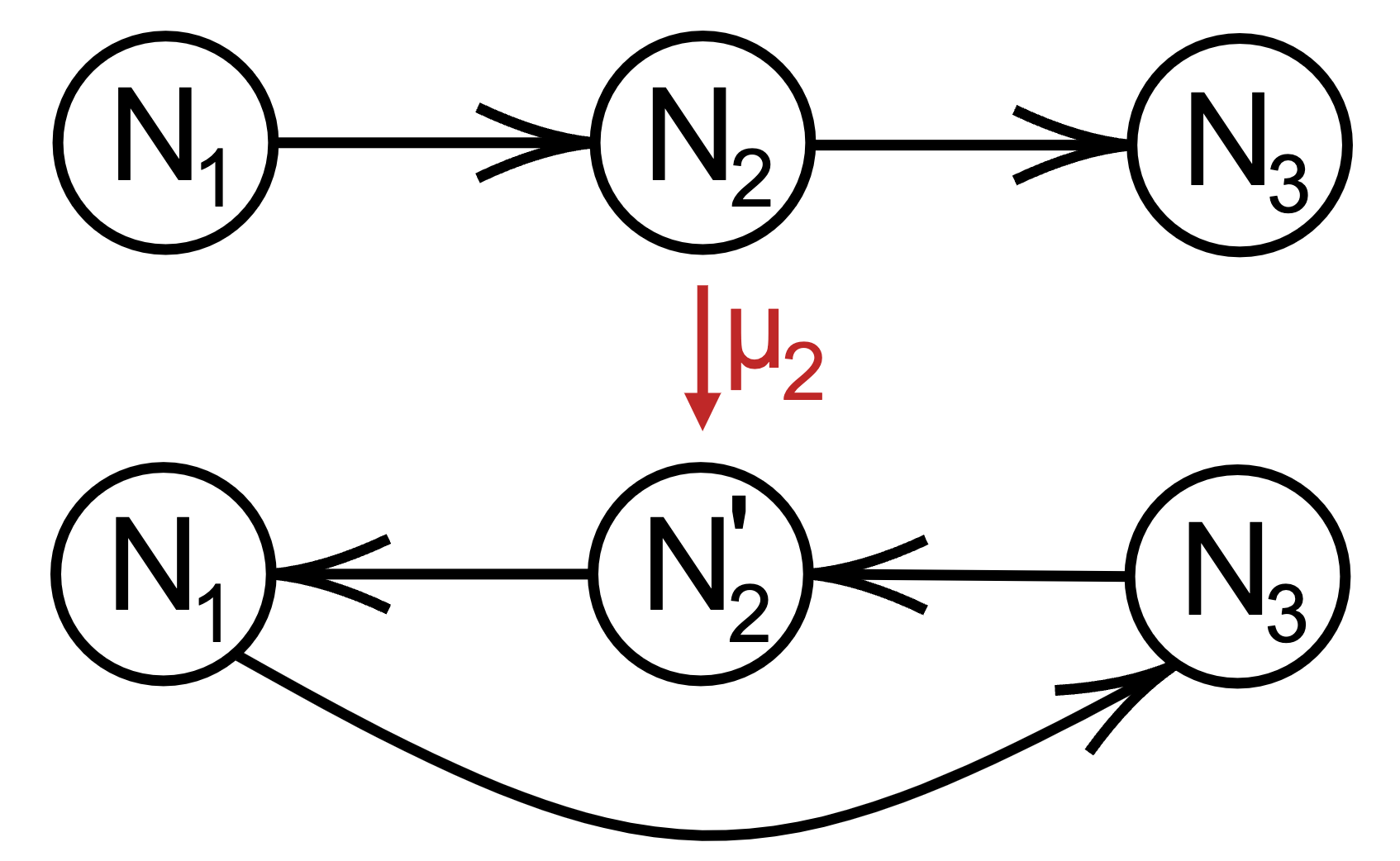}
    \caption{Mutation of an $\mathcal{N}=1$ quiver at the central node to produce an infra-red equivalent theory, under the quiver-interpreted rules of Seiberg duality. The node's rank updates $N_2 \longmapsto N_2' = N_1 - N_2$.}
    \label{quivermutation_example}
\end{figure}

Quiver mutation was first discovered by Seiberg and hence also goes by the name of Seiberg duality \cite{Seiberg:1994pq}. Given a quiver the process to generate an equivalent quiver follows a simple series of steps:
\begin{enumerate}
    \item Select a node of the quiver.
    \item Reverse the direction of all edges incident to that node.
    \item Where there exists a 2-path passing through the selected node, introduce an edge to extend this into a 3-cycle.
    \item Where introduction of edges leads to any 2-cycles, remove the 2-cycles.
    \item Update the selected node's gauge group rank by subtracting the original rank from the sum of the ranks of the gauge groups at the tail ends of any edges the selected node was at the head end of prior to mutation.
\end{enumerate}
The process of quiver mutation is shown explicitly for an example quiver in Figure \ref{quivermutation_example}. 

Performing quiver mutation on any node of the quiver will produce an infra-red equivalent quiver. Hence, all quivers produced from mutation of a given quiver at each of its nodes are infra-red equivalent, and by extension any mutation of those quivers are equivalent also. Iteration of the quiver mutation process then leads to an entire class of equivalent quiver gauge theories, grouped into a duality tree \cite{Klebanov:2000hb,Franco:2003ja}. For quiver mutation there are importantly two mutation types: finite-mutation and infinite. For finite-mutation type quivers exhaustive mutation of all nodes iteratively will only ever generated a finite number for quivers, whilst for infinite type there will be some sequences of mutation which will continue to generate new quivers.

While generating new quivers that are equivalent is relatively easy via quiver mutation, determining whether or not any two given quivers are equivalent can be very difficult. This leads one to wonder whether modern ML methods would be able to detect dualities between quivers. This was the focus of the work in \cite{Bao:2020nbi}, the results of which we will discuss later\footnote{The mathematical process of mutation has roots in another field of mathematics, known as cluster algebras \cite{CA_1}. The application of ML techniques and network science analysis to cluster algebra mutation was initiated, in the papers \cite{Dechant:2022ccf,Cheung:2022itk}.}.

\subsection{Brane Webs}\label{sec:brane_webs}

\begin{figure}[!h]
    \centering
    \begin{subfigure}{0.49\textwidth}
        \centering
        \includegraphics[width=\textwidth]{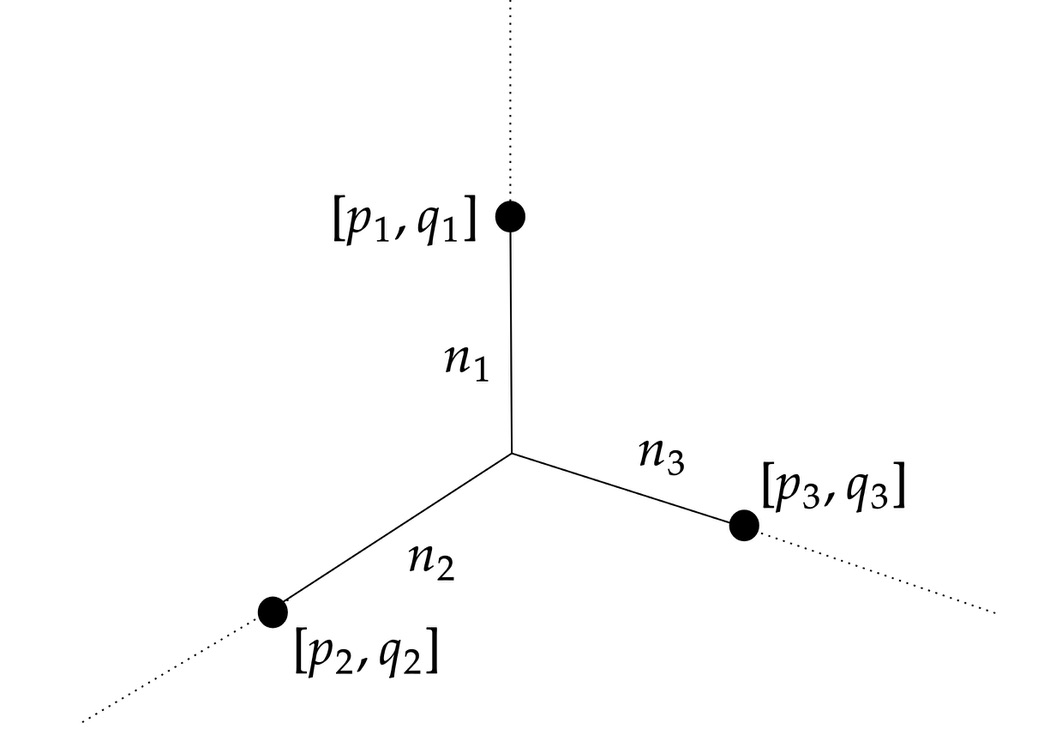}
        \caption{}\label{GeneralWeb}
    \end{subfigure}
    \begin{subfigure}{0.49\textwidth}
        \centering
        \includegraphics[width=0.8\textwidth]{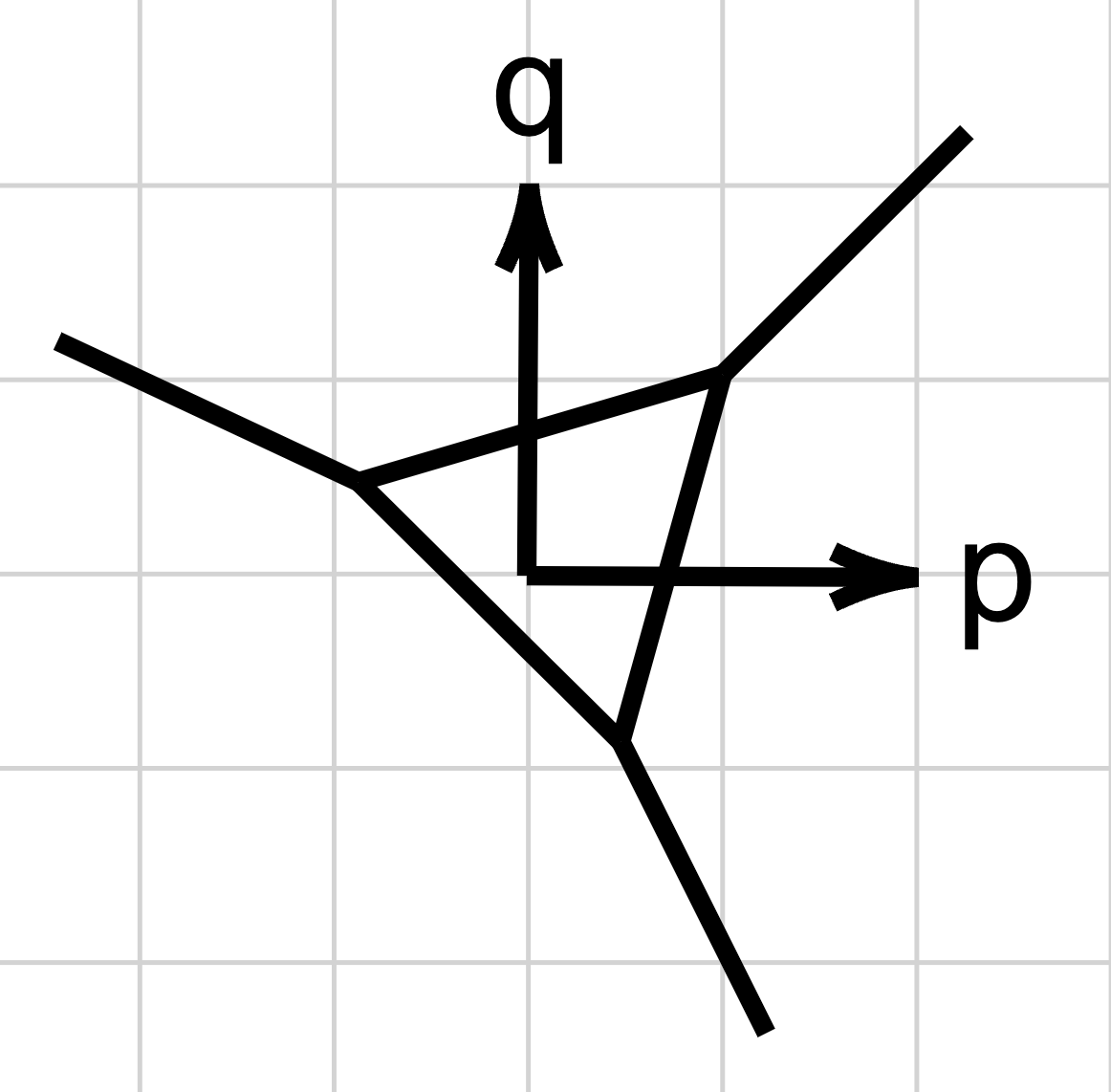}
        \caption{}\label{ExampleWeb}
    \end{subfigure}
    \caption{Three-leg $(p,q)$ 5-brane webs, for (a) the general case, and (b) the equivalent web for the trinian quiver theory with moduli space as the lattice polygon with vertices $\{(1,0),(0,1),(-1,-1)\}$.}
    \label{braneweb_example}
\end{figure}

As we mentioned earlier, string theory relaxes the assumption that the fundamental building blocks of the universe are 0-dimensional particles, and introduces 1-dimensional strings. In fact, there is no need to stop at 1-dimension, we can consider building blocks of 2, 3 or even higher dimensions. This is because in string theory we are dealing with 9 spatial dimensions rather than 3. We refer to such $p$-dimensional objects as `$p$-branes', and so in this terminology a string is also called a 1-brane. In Type IIB (one of the five versions of string theory), an NS5-brane is a 5-brane that is magnetically charged under the Kalb-Ramond field and the D5-brane is a 5-brane that is magnetically charged under the Ramond-Ramond field. A $(p,q)$ 5-brane is then a bound state of $p\in\mathbb{Z}$ D5-branes and $q\in\mathbb{Z}$ NS5-branes. Several $(p,q)$ 5-branes are allowed to meet at a point so long as magnetic charge is conserved, i.e., $\sum_{i}p_{i} = \sum_{i}q_{i} = 0$. We call the system of three or more such 5-branes meeting at a point, a 5-brane web. These brane webs can be represented by a simple diagram where the $(p,q)$ 5-branes are represented by line segments, as shown in Figure \ref{braneweb_example}. These brane webs are interesting to study because they describe certain quantum field theories at the junction. 

We can introduce $[p,q]$ 7-branes on which the $(p,q)$ 5-branes end, these are shown as dots in Figure \ref{braneweb_example}. Furthermore, there can be more than one $(p,q)$ 5-brane `hanging' from a $[p,q]$ 7-brane and therefore we must introduce further variables $n_{i}$ that represent the number of 5-branes in each leg. All the information of a 5-brane web with $N$ legs can then be encoded in a $2\times N$ web matrix
\begin{equation}\label{eq:web_matrix}
    W=\begin{pmatrix}
        n_{1}p_{1} & n_{2}p_{2} & \cdots & n_{N}p_{N} \\
        n_{1}q_{1} & n_{2}q_{2} & \cdots & n_{N}q_{N}
    \end{pmatrix}
\end{equation}

An important feature of 7-branes is that they have an $SL(2,\mathbb{Z})$ monodromy associated to them, which is accounted for by the branch cut (dashed line) in the brane web diagram. This means that when one 7-brane crosses the branch cut of another 7-brane, the $[p,q]$ charges of the 7-brane crossing the cut are changed by the monodromy of the other 7-brane. The monodromy matrix of a $[p,q]$ 7-brane is given as
\begin{equation}\label{eq:monodromy}
    M_{[p,q]} = 
    \begin{pmatrix}
    1 + pq & -p^{2} \\
    q^{2} & 1 - pq 
    \end{pmatrix}.
\end{equation}
The position of the branch cut is irrelevant and therefore one can position the cut of all 7-branes so that they go away from the brane web, such that no 5-branes cross them. The introduction of 7-branes also introduces further constraints on brane webs. Namely, the self-intersection number $\mathcal{I}$ of the junction must satisfy the constraint $\mathcal{I}\geq-2$, where $\mathcal{I}$ is defined as 
\begin{equation}\label{eq:self_intersection}
\mathcal{I} = \left| \sum_{1 \leq i < j \leq L} \text{det} \begin{pmatrix}
p_{i} & p_{j} \\
q_{i} & q_{j} 
\end{pmatrix} \right| - \sum_{i=1}^{L} [\text{gcd}(p_{i},q_{i})]^{2} 
\end{equation}

The physics described by a brane web is not unique, in other words, there exists equivalence relations between webs such that two webs may appear different but in fact describe the physics. In \cite{dewolfe1998} it was conjectured that equivalent sets of 7-branes, which are equivalent to brane webs in the three-leg case, are completely determined by three invariants: $\mathcal{I}$, $M_{tot}$ and $\ell$. Here $\mathcal{I}$ is the self-intersection number defined in \eqref{eq:self_intersection}, $M_{tot}$ is the total monodromy
\begin{equation}\label{eq:total_monodromy}
    M_{tot} = M_{(p_{1},q_{1})}M_{(p_{2},q_{2})}M_{(p_{3},q_{3})},
\end{equation}
where $M_{(p_{i},q_{i})}$ is the monodromy of the $i$-th leg defined by \eqref{eq:monodromy}, and $\ell$ is the asymptotic charge
\begin{equation}\label{eq:asymptotic_charge}
    \ell = \text{gcd} \biggl\{ \text{det}
    \begin{pmatrix}
        p_{i} & p_{j} \\
        q_{i} & q_{j}
    \end{pmatrix}
    \forall i,j \biggr\}.
\end{equation}
It turns out that these three invariants are not enough to completely determine equivalence between brane webs, but still provide a necessary condition. The authors in \cite{Arias-Tamargo:2022qgb} trained a neural network to determine whether or not two three-leg 5-brane webs are ``weakly equivalent'', meaning that they share the same $\mathcal{I}$, $M_{tot}$ and $\ell$. We will present the results of this paper in the following sections.

\section{Supervised Learning}\label{sec:supervised} 

\subsection{Neural Networks}\label{sec:nn}

To explain how neural networks work, we begin by introducing the building block of any neural network; a neuron. A neuron takes in a set of input data $\{x_{i}\}$ and produces a single numerical output $\hat{y}$. We can divide the function of the neuron into three parts:
\begin{enumerate}
    \item Firstly, each input $x_{i}$ is multiplied by a weight $w_{i}$: $w_{i}x_{i}$.
    \item Next, all the weighted inputs are summed and a bias $b$ is added: $\sum_{i}w_{i}x_{i}+b$.
    \item Finally, the sum is passed through an activation function which produces an output: $\hat{y}=f(\sum_{i} w_{i}x_{i}+b)$.
\end{enumerate}
A neural network is simply a collection of neurons connected together in a series of layers. In between the input and output layers are one or more `hidden' layers. 

\paragraph{Training}\mbox{}\\
The process of training a neural network involves repeatedly calculating the `error' of the model outputs so that the weights and biases can be updated in order to reduce the error. Computing the error requires a choice of loss function. Typical loss functions for regression problems are mean absolute error (MAE)
\begin{equation}\label{eq:MAE}
    MAE = \frac{1}{n} \sum_{i=1}^{n} |y_{i} - \hat{y}_{i}|,
\end{equation}
and mean square error (MSE)
\begin{equation}\label{eq:MSE}
    MSE = \frac{1}{n} \sum_{i=1}^{n} (y_{i} - \hat{y}_{i})^{2},
\end{equation}
where $y_{i}$ and $\hat{y}_{i}$ are the true and predicted values respectively. For classification problems the default loss function is cross entropy (CE)
\begin{equation}
    CE = -\sum_{i=1}^{n}y_{i}\log{(\hat{y}_{i})},
\end{equation}
for multiclass classification, and binary cross entropy (BCE)
\begin{equation}\label{eq:BCE}
    BCE = -\sum_{i=1}^{n}(y_{i}\log{(\hat{y}_{i})} + (1-y_{i}), \log{(1-\hat{y}_{i})})
\end{equation}
for binary classification.

The method by which we change the weights and biases to minimise the loss is called the optimisation algorithm. The simplest optimisation algorithm is called gradient descent, which determines which direction the model parameters should be altered so that the loss function can reach the minima. In the gradient decent optimisation algorithm, the model parameters $\theta$ are adjusted as follows:
\begin{equation}
    \theta_{t+1} = \theta_{t} - \alpha \frac{\partial L}{\partial \theta}
\end{equation}
where $\alpha$ is a parameter called the learning rate that controls the speed and accuracy of training. In mini-batch gradient descent, the training dataset is divided into smaller batches, and the model parameters are updated after every batch. 

Gradient descent methods of optimisation present some challenges. Firstly, choosing a value for the learning rate can be difficult. A value that is too small can lead to painfully slow convergence. On the other hand, a value too large can cause the loss function to overshoot the minimum and in some cases diverge. Another problem that arises in gradient descent methods is getting trapped at local minima. There exist more advanced optimisation methods, such as Adam \cite{kingma2017}, that tackle these challenges but we wont explain these here.

\paragraph{Evaluation}\mbox{}\\
For regression tasks, typical performance metrics include MAE \eqref{eq:MAE}, MSE \eqref{eq:MSE} and the $R^{2}$ score, which is defined as the proportion of the variance in the dependent variable that is predictable from the independent variable(s). Therefore, an $R^{2}$ score close to 1 means the regression model is a good fit, whereas a score close to 0 means the model is a poor fit. The equation for computing $R^{2}$ is as follows
\begin{equation}
    R^{2} = 1 - \frac{\sum_{i=1}^{n}(y_{i}-\hat{y}_{i})^{2}}{\sum_{i=1}^{n}(y_{i}-\bar{y})^{2}},
\end{equation}
where 
\begin{equation}
    \bar{y} = \frac{1}{n} \sum_{i=1}^{n} y_{i}.
\end{equation}

For classification tasks, the default performance metric is accuracy 
\begin{equation}\label{eq:acc}
    acc = \frac{TP+TN}{TP+FP+TN+FN},
\end{equation}
where $TP,FP,TN,FN$ denote the true positive, false positive, true negative, and false negative counts respectively. Accuracy, however, can sometimes be misleading. For example, consider the binary classification situation where the first class occurs 99\% of the time and the second class occurs only 1\% of the time. An algorithm that just predicts the first class every time would give an accuracy score of 99\% which looks great but this isn't actually a good model. A better metric in this case is Mathew's correlation coefficient (MCC)
\begin{equation}\label{eq:MCC}
    MCC = \frac{TP \times TN - FP \times FN}{\sqrt{(TP+FP)\times(TP+FN)\times(TN+FP)\times(TN+FN)}},
\end{equation}
which is similar to the Pearson correlation coefficient, where a score of 1 indicates complete agreement between predictions and truth, 0 indicates that the predictions are no better than random guessing, and -1 indicates complete disagreement between the predicted and truth. MCC is a better choice of metric when dealing with classes of different sizes. 

In order to obtain an unbiased evaluation of a neural network, one shuffles and then splits the data into two sets: a training set and a test set. The usual train:test split is 80:20. The training dataset is used the fit the model and  then the trained model is used to make predictions on the unseen test set. Cross-validation is another method commonly used to get an unbiased evaluation, whereby the data is shuffled and then split into $k$ groups, then each group acts as the test group once and the remaining groups are combined to create the training set. Each time the model is trained on the training set, evaluated on the test set and the evaluation scores are recorded. The mean and and standard deviation of the evaluation scores are then calculated and used to measure the model performance.

\paragraph{Example: Polytope Properties}\mbox{}\\ 
As introduced in §\ref{sec:polytopes}, lattice polytopes can be used to construct CY manifolds. The geometric properties of the resulting CY can be extracted directly from the polytope and its dual. How these properties can be learnt directly from the input polytopes is probed through the lens of ML in the work of \cite{Bao:2021ofk}.
Here focus was put on 2 and 3-dimensional polytopes\footnote{ML analysis of the 4-dimensional reflexive polytopes of the Kreuzer-Skarke dataset \cite{Kreuzer:2000xy,Altman:2014bfa} is available at \cite{Berglund:2021ztg}.}, using the database \cite{Kas08}. A selection of simple properties were investigated, with the aim that future work would develop the application to more subtle properties\footnote{Extension may also consider the non-compact toric variety construction as performed in \cite{Krefl:2017yox} using lattice polytope toric diagrams.}.

The properties focused on included: volume, dual volume, reflexivity, Gorenstein index, and codimension, each having a simple interpretation in terms of the input polytope. The volume is simply the volume of the polytope in the lattice with an appropriate dimensional normalisation to ensure integer outputs, whilst the dual volume is the same volume but for the dual polytope of the input. Gorenstein index is a measure associated to reflexivity, and is the minimum integer scaling of the dual polytope required such that it becomes lattice. Finally the codimension dictates the difference in dimension between the Fano variety and the projective variety it is embedded within. Each of these properties relies on non-trivial formulas for computation, particularly those relying on the computation of the respective dual. For example, the codimension requires computing the Hilbert basis cardinality of cone of the dual polytope, which is associated to the number of lattice points in and on the dual polytope. We also note that due to the sparsity of reflexivity of 2-dimensional lattice polygons (only 16 unique reflexive polygons up to transformation), reflexivity was only considered for 3-dimensional polyhedra.

The first key result of \cite{Bao:2021ofk} arose from the subtleties in polytope representation.  The standard way to represent a polytope is the vertex list, however as hinted to previously there are many vertex combinations that give the same polytope and ultimately the same toric geometry. An example of this is rotating a polytope through $\pi$ radians about one of the lattice coordinate axes, the polytope is ultimately the same despite now having different vertices. To circumvent this representation redundancy the Pl\"ucker coordinates of the polytope were used as input. These coordinates are $GL(n,\mathbb{Z})$-invariant, and when only considering polytopes whose vertices generate the integer lattice this representation becomes unique. The Pl\"ucker coordinates are the maximal minors of the vertex matrix kernel, and although they do remove the $GL(2,\mathbb{Z})$ redundancy they still express the permutation redundancy in the coordinate ordering -- which was used for data augmentation.

Use of the Pl\"ucker coordinates substantially improved learning results, in some cases reducing the MAE measure by over an order of magnitude. To exemplify the Pl\"ucker coordinate computation we return to the $\{(1,0),(0,1),(-1,-1)\}$ polygon example. This vertex matrix has kernel: $(1,1,1)$ from adding the three vertices' coordinates, and since the kernel matrix has dimension $1 \times 3$ the maximal minors are simply the $1 \times 1$ entries: $(1,1,1)$.

Training the dense NNs used to learn each of the considered properties from Pl\"ucker coordinate input lead to a mix of results.
For both the 2 and 3-dimensional datasets learning the volume and dual volumes lead to MAE scores of order $\sim 1$, which was surprisingly strong considering the 2-dimensional volume range was $\sim 500$. These result imply the existence of simple formulas for the volume and dual volumes from the Pl\"ucker coordinates left to be found in future work. This existence may be intuitively expected due to the common determinant-nature of volume and minor computations. Reflexivity binary classification for 3-dimensional polyhedra performed at accuracies $\sim 0.8$; whilst for the 2-dimensional polygons learning the Gorenstein index had MAE scores $\sim 5$ and for codimension $\sim 2$ relative to ranges of 29 and 40 respectively.

This disparate scores illustrate the relative functional complexity for computing each of these properties. Gorenstein index relies on integer truncation which the NNs struggle with, while the many steps of the codimension computation seem to be learnt with some non-trivial success. How these learnt NN functions can be interpreted for mathematical insight is an prudent endeavour for future work.

\paragraph{Example: Amoebae Coefficients}\mbox{}\\ 
The amoeba projection of complex polynomials, as discussed in §\ref{sec_amb}, does not necessarily preserve the polynomial's topology. Where the geometry defined by a polynomial in $\mathbb{C}^n$ has genus $g$, the respective amoeba will have genus $\in \{0,1,2,...,g\}$ dependent on the coefficient choices in the defining polynomial.

The aim of the work in \cite{Bao:2021olg,Chen:2022jwd} involves the use of a NN architecture to learn the genus of the respective amoeba given the vector of coefficients for the polynomial. The classification over the range of genus values up to $g$ was learnt with accuracies up to $0.99$ and $0.94$ for the simpler $(\mathbb{P}^1)^{\times n}$ examples respectively in $n=2,3$ dimensions. Furthermore, where the set the coefficients were sampled from was less restrictive, the learning performance lowered.
For more complicated examples, where the $g$ value was higher, the accuracies dropped to around $\sim 0.9$, still showing strong performance.

These strong results then motivated the use of interpretable ML techniques to extract the bounds in the coefficient space where the amoeba genus changed, with satisfactory successes.

\subsubsection{CNN} 

Convolutional neural networks (CNNs) are a specific type of NN, where the connections between layers are restricted to connect neurons to those adjacent to the equivalent neuron in the next layer. 
This way the architecture is explicitly forced to focus on local structure, which in turn has lead to significant advances in the field of image recognition.

As noted, the architecture is well suited for handling higher-dimensional data structures, such as images. 
Hence, where the dimensionality of the objects representation is important the use of CNNs may be well motivated.
This motivation also applies for graph-theoretic data which is simply represented with the respective adjacency matrices, although this representation is in some sense distinctly non-local due to graph isomorphism.
The 2-dimensional nature of complete intersection configuration matrices has also made this architecture style more desirable when analysing this type of CY manifolds \cite{He:2017aed,Bull:2018uow,Bull:2019cij}.

\paragraph*{Example: Quiver Mutation}\mbox{}\\ 
With the application to adjacency matrices in mind, we return to the question of determining whether two given quivers are equivalent.
In standard practise one would have to exhaustively compute the mutations of one of the quivers until the other was produced to confirm their equivalence. However as some duality trees are infinite one could never exhaustively prove theories were inequivalent.
To attack this seemingly infeasible problem the work of \cite{Bao:2020nbi} initiates the implementation of tools from ML, more specifically CNNs\footnote{The work in \cite{Bao:2020nbi} dictates learning results with both the Naive Bayes and CNN architectures, both performed comparatively so are quoted as CNN results here.}, to learn how to identify equivalent theories.

In investigating the effectiveness of the ML architectures in learning whether a given two quivers are equivalent, the learning performance was also compared against the mutation type of quivers given.
In performing the investigations a sample of quivers from each type were taken, each quiver with a small number of nodes such that mutation to a high depth was feasible.
Importantly the finite-mutation type quivers included those constructed from adding orientations to the edges of simply-laced Dynkin diagrams, which are naturally all of finite-mutation type. 

Three primary investigations were carried out, each set up as a classification task for the respective ML architectures.
In each case the architecture received the quivers as adjacency matrix data.

The first investigation sampled pairs of quivers from a database of quivers formed from the combination of two duality trees.
Each duality tree was generated until an approximate number of 500 quivers were produced, then approximately 20,000 pairs were sampled from all possible choices -- ensuring a balanced dataset of pairs being equivalent and not equivalent.
The ML architecture then performed binary classification on the pairs to determine whether the two quivers were equivalent.
For all combinations of types for the trees: both finite-mutation, one finite-mutation and one infinite, and both infinite; the classification was perfect with accuracy and MCC scores of 1.
Validating clearly that the architecture can identify invariants under the mutation and learn how to identify quivers describing the same infra-red theory.

The second investigation developed the previous line of inquiry to test the ML architecture for binary classification on pairs now sampled from a database of quivers from $>2$ duality trees.
Where three trees were used the performance measures dropped to accuracy $\sim 0.9$ and MCC $\sim 0.8$ for all tested combinations of types, and where more trees were used the performance further dropped down to a tested performance of 0.75 accuracy on a database of 6 trees.
This implies that whilst the architectures can learn the equivalence under mutation this becomes more difficult with more varied and sparser data, as perhaps expected from standard data science intuition.

The third and final investigation performed the multiclassification directly. In the direct multiclassification the CNN would receive a adjacency matrix of a single quiver and have to assign it a label attributed to the duality tree it came from.
The performance reached accuracies and MCCs $\sim 0.9$ when classifying between three finite-mutation type trees, and between a set of three trees where two were finite-mutation and the other infinite. However performance was not strong between three infinite type trees, dropping to $\sim 0.4$ performance measures.

Further investigation extensions included enhancing the data with the gauge group rank information of the quiver gauge theory which marginally improved performances; extrapolating prediction to quivers many mutations away from the low-mutation depth set trained on which performed well with accuracies $\sim 0.75$; and 
identifying quivers from a set including random antisymmetric matrices with accuracies up to 1.

\paragraph{Example: Amoebae Images}\mbox{}\\ 
Practically, the analytic computation of an amoeba is computationally expensive.
To bypass this they are often visualised through a Monte Carlo sampling, whereby points on the complex polynomial are sampled and projected onto the real space.
However, due to the non-linearity of the projection, creating a uniformly sampled amoeba image is exceptionally difficult.
This easily leads to erroneous identification of the amoeba genus, where certain parts of the amoeba body are poorly sampled; and with a desire to identify the genus for a large dataset of coefficients, computational techniques and ML become a sensible route for streamlining the process.

In the work of \cite{Bao:2021olg}, CNNs are used to learn the genus from the Monte Carlo generated images directly, avoiding the need for expensive analytic computation.
Focusing on the prototypical $\mathbb{P}^1 \times \mathbb{P}^1$ surface, leading to amoebae with genus values at most 1, the CNN binary classifiers could reach accuracies up to $0.99$ and MCC scores up to $0.97$.
In performing the classification a coarse-graining of the filter was first applied to the images, in the spirit of the pooling kernel layers typical for CNN architectures, and the resolution reduction interestingly had an optimum value where local structure was sufficiently averaged to smooth out the non-linear sampling.

\subsubsection{SNN}\label{sec:snn}

Siamese Neural Networks (SNNs), first introduced in \cite{bromley1993} in the context of signature verification, are neural network architectures consisting of two or more identical sub-networks that determine the similarity of inputs. The sub-networks $f_{w}$ map elements of a dataset $\mathcal{D}$ to points in $\mathbb{R}^{d}$, where $w$ denotes the weights and biases of the network. The goal is to train the network such that after training similar elements of $\mathcal{D}$ are mapped close together in $\mathbb{R}^{d}$, and dissimilar elements are mapped far apart. The weights and biases are determined by extremising a loss function that depends on the squared Euclidean distance between the embeddings $||f_{w}(W_{1}) - f_{w}(W_{2})||^{2}$.

SNNs are commonly trained using the triplet loss, which takes an anchor (A), positive (P) and negative (N) input triple, where the anchor-positive inputs are similar and the anchor-negative inputs are dissimilar. If the network is performing well, the distance between the anchor and positive embeddings will be smaller than the distance between the anchor and negative embeddings:
\begin{equation}
    ||f_{w}(A)-f_{w}(P)||^{2} \leq ||f_{w}(A)-f_{w}(N)||^{2}.
\end{equation}
Equivalently, we could write
\begin{equation}
    ||f_{w}(A)-f_{w}(P)||^{2} - ||f_{w}(A)-f_{w}(N)||^{2} \leq 0.
\end{equation}
This inequality is satisfied if the $f_{w}$ always outputs a constant value, say 0. To prevent the network from learning this, we modify the objective by introducing a hyperparameter $\alpha \in \mathbb{R}_{>0}$:
\begin{equation}
    ||f_{w}(A)-f_{w}(P)||^{2} - ||f_{w}(A)-f_{w}(N)||^{2} + \alpha \leq 0
\end{equation}
The triplet loss function is then defined as:
\begin{equation}\label{eq:triplet_loss}
    \mathcal{L} = \max{(||f_{w}(A)-f_{w}(P)||^{2} - ||f_{w}(A)-f_{w}(N)||^{2} + \alpha, 0)}
\end{equation}
This is the loss of a single triplet, the overall loss function of the network is then the average of these individual losses over a batch of $\mu$ triplets:
\begin{equation}
    J = \frac{\sum_{i=1}^{\mu}\mathcal{L}(f_{w}(A^{(i)}),f_{w}(P^{(i)}),f_{w}(N^{(i)}))}{\mu}
\end{equation}

\begin{figure}[h]
    \centering
    \includegraphics[width=10cm]{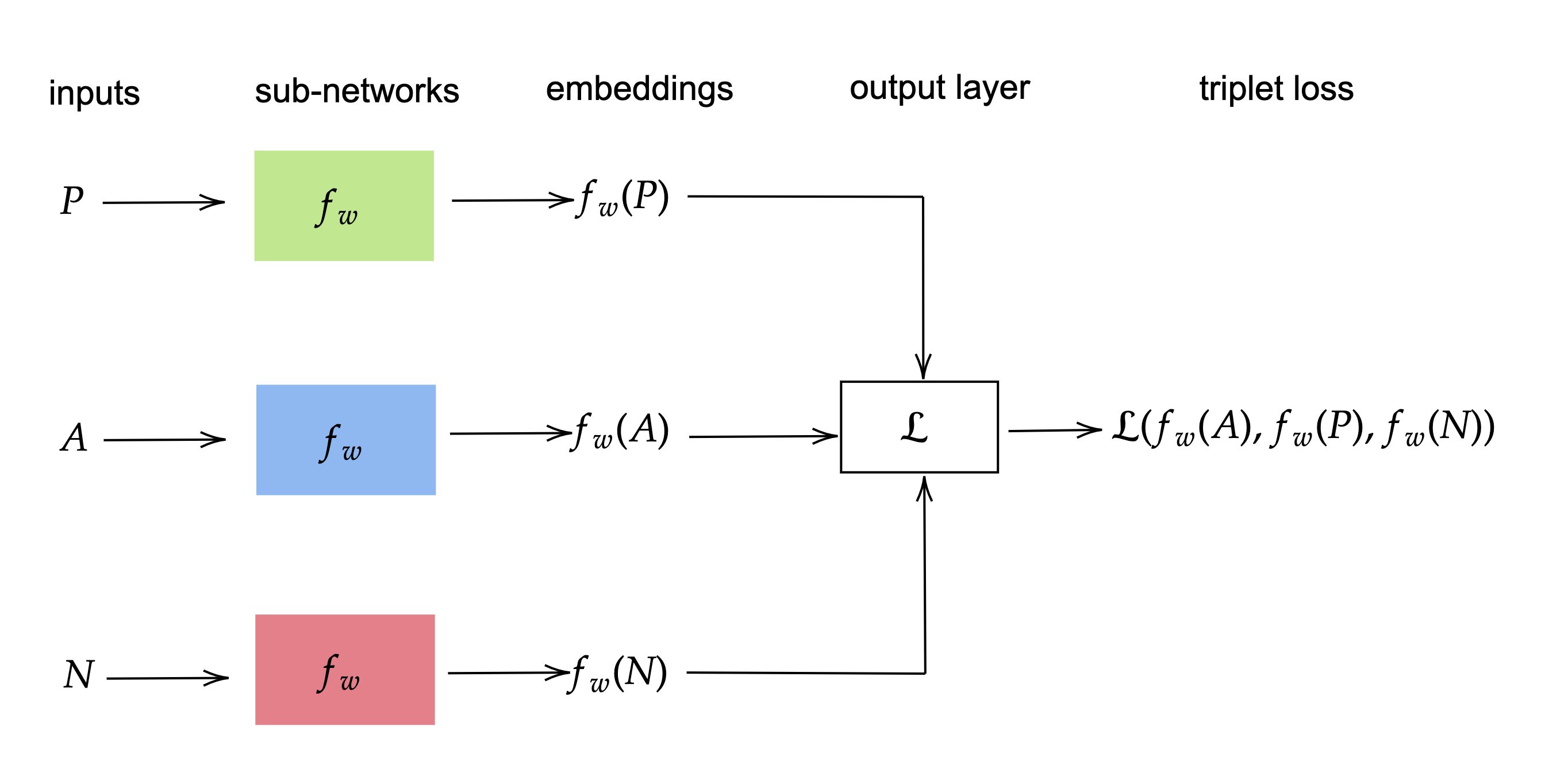}
    \caption{Siamese neural network diagram using triplet loss.}
    \label{fig:SNN}
\end{figure}

Using the triplet loss function, a network is made up of three identical sub-networks that produce three embeddings for the anchor, positive and negative inputs. The outputs of these sub-networks are fed to an external output layer that computes the triplet loss. Figure \ref{fig:SNN} show the architecture of a typical SNN trained with triplet loss. 

\paragraph{Example: Brane Webs}\mbox{}\\ 
It is difficult to determine by hand whether two 5-brane webs are equivalent or not. It is therefore an interesting question to ask whether a neural network could determine whether two brane webs are equivalent or not. This is exactly what the authors in \cite{Arias-Tamargo:2022qgb} tested for three leg brane webs using an SNN.  

An SNN was trained, using triplet loss, to determine whether two webs are equivalent, where the input data was in the format of $2\times3$ web matrices \eqref{eq:web_matrix}. After training on triplets taken from the training set, the trained subnetwork was then used to produce an embedding for every web in the remaining test set. Two webs taken from the test set were predicted to be equivalent if the squared Euclidean distance between their embeddings was less than some threshold value and inequivalent otherwise. The result of these pairwise equivalence predictions was that the SNN was able to detect equivalence between webs taken from two equivalence classes with 100\% accuracy.

\subsection{Support Vector Machines}\label{sec:svm} 

Support vector machines (SVMs) are simple architectures designed specifically for binary classification of high-dimensional data in $\mathbb{R}^n$. The SVM seeks to optimise the position of an $(n-1)$-dimensional linear hyperplane in $\mathbb{R}^n$ which best separates the two classes. The linearity requirement for the hyperplane can be dropped through the use of kernel methods, which introduce non-linear terms into the hyperplane equation by embedding the data in a $\tilde{n}$-dimensional space (for $\tilde{n}>n)$ where the extra dimensions correspond to nonlinear terms, then performing the linear hyperplane optimisation there and projecting back down to the original $n$-dimensions.

Since an explicit hyperplane is generated through the SVM algorithm this hyperplane has the potential for direct interpretation in the physical context of the mathematical data; as well as simple interpretable extrapolation of the classification to objects beyond the sampling range.

\paragraph{Example: Calabi-Yau Threefolds}\mbox{}\\ 
The Calabi-Yau threefold database constructed from hypersurfaces in $\mathbb{P}^4$ is represented by a set of 7555 5-vectors of weights \cite{Candelas:1989hd}. 
In the work of \cite{Berman:2021mcw}, to probe the importance of each necessary condition on these 5-vectors, conjugate datasets which did not satisfy each condition were generated.

A selection of supervised ML architectures were then used to binary classify whether a given 5-vector of weights (and hence a given 4-dimensional weighted projective space) admitted a CY threefold hypersurface. Despite NNs performing the best, simpler architectures were concentrated on due to their interpretability. Notably including the use of SVMs.

SVMs could differentiate 5-vectors which admitted a CY hypersurface with accuracies up to $0.75$. The consistency across the binary classification of the CY dataset against each of datasets with progressively more necessary conditions included indicated that the CY property occurred more subtly in the 5-vectors than for the properties considered.

Further investigation of the correlation between the SVM misclassifications and the CY Hodge numbers, indicated which of the non-CY datasets used for training made the classifier better over different ranges of these topological Hodge numbers.
In particular, if the transverse property\footnote{For the 5-vector's weights $w_i$ the transversity property implies that $\forall w_i \ \exists w_j \ s.t. \ \frac{\sum_k(w_k) - w_j}{w_i} \in \mathbb{Z}^+$.} was not included in the non-CY dataset the classifiers performed perfectly across the high $h^{2,1}$ range, while if it was included they instead only performed perfectly across the high $h^{1,1}$ range.
These unexpected correlations indicate the intimate link between transversity of a weighted projective space and the Hodge number of the subsequent hypersurfaces. A relationship that would be interesting to solidify in future work.

\section{Unsupervised Learning}\label{sec:unsupervised} 

The program of mathematical research often follows the paradigm of conjecture formulation followed by proof.
The first step of conjecture formulation involves testing example cases of the examined mathematical objects to identify phenomena which they appear to obey.
However to be confident in a conjecture a large number of examples must be confirmed for sensible extrapolation of the conjectured relation.
This step lends itself naturally to techniques from big data, not only for the processing of large sets of examples, but also for the identification of the patterns which lead to the conjectures.

Unsupervised learning includes a large class of techniques, perhaps most importantly including methods for processing datasets to extract patterns and fundamental degrees of freedom.
The unsupervised subfield of machine learning can hence be further subdivided into techniques for feature extraction which identify the most significant and relevant parts of the data, and techniques for clustering which appropriately groups the data and assesses similarities and symmetries.

Each of these classes of methods have spirit manifestly derived from the scientific method in physics. 
To truly understand the universe we live in we require efficient prediction, which relies on knowledge of the relative importance's of each component in a calculation -- naturally associating to feature extraction. 
Whilst to most appropriately develop intuition for extension of theorised models an understanding of the theory's symmetries helps reduce the amount of work required -- naturally extractable from clustering methods.

In this section, we discuss how some of the fundamental techniques within unsupervised learning are applied to selected problems in high-energy theory. The feature extraction methods include principal component analysis (PCA); t-distributed stochastic neighbour embedding (t-SNE); and topological data analysis (TDA); all useful for analysis of the datasets of mathematical objects and their visualisation.
Additionally, the application of the prototypical clustering method, $k$-Means clustering, is discussed to aid with classification of the respective objects.

\subsection{Principal Component Analysis}\label{sec:pca} 

Approximation is an integral part of computation in physics.
Crucial to accurate approximation is the identification of the importance of each contributing term to a calculation. Principal component analysis, or PCA, provides a method to quantify the relative importance of data features. Specifically, for a vector representation it identifies the linear combinations of the vectors' entries which best describe the data's variation.

Also important for data representation is the independence of the components used, a property that is implicitly enforced in the PCA method of covariance matrix diagonalisation. Independence of components ensures that the components used are truly fundamental, and that the physical representation is maximally efficient for the simplicity of representation.
This effect is ubiquitous in physical theory, perhaps most commonly introduced through the normal mode representation for string vibration.

PCA therefore provides a method of linear feature extraction of objects represented by vectors, and the linearity restriction can be further generalised to non-linear combinations with kernel methods akin to those used for SVMs; as discussed in §\ref{sec:svm}.

\paragraph{Example: Calabi-Yau Threefolds}\mbox{}\\ 
The weighted $\mathbb{P}^4$ database of 5-vectors admitting a CY hypersurface of \cite{Candelas:1989hd}, analysed in \cite{Berman:2021mcw}, is too high-dimensional for direct visualisation.
Since the 5-vector's weights have no implicit logical order, they are naturally listed in ascending order. 
This leads to the most dominant components for the variation naturally being the later entries.
This behaviour was reflected in the PCA decomposition of the datasets considered.

\begin{figure}[tb]
	\centering
	\begin{subfigure}{0.45\textwidth}
		\centering
		\includegraphics[width=\textwidth]{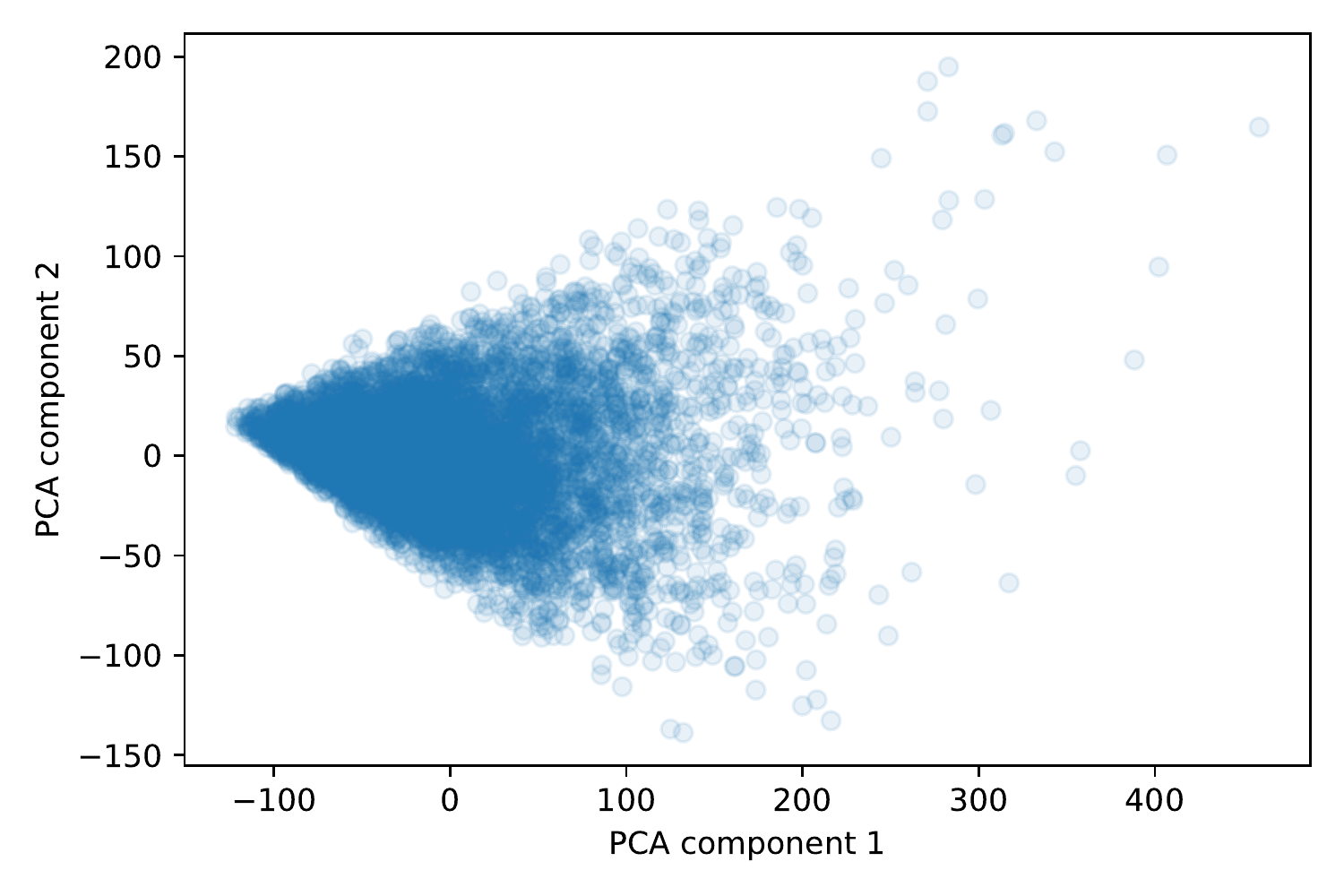}
		\caption{Random 5-vectors}\label{RandomPCA}
	\end{subfigure} 
    \begin{subfigure}{0.45\textwidth}
    	\centering
    	\includegraphics[width=\textwidth]{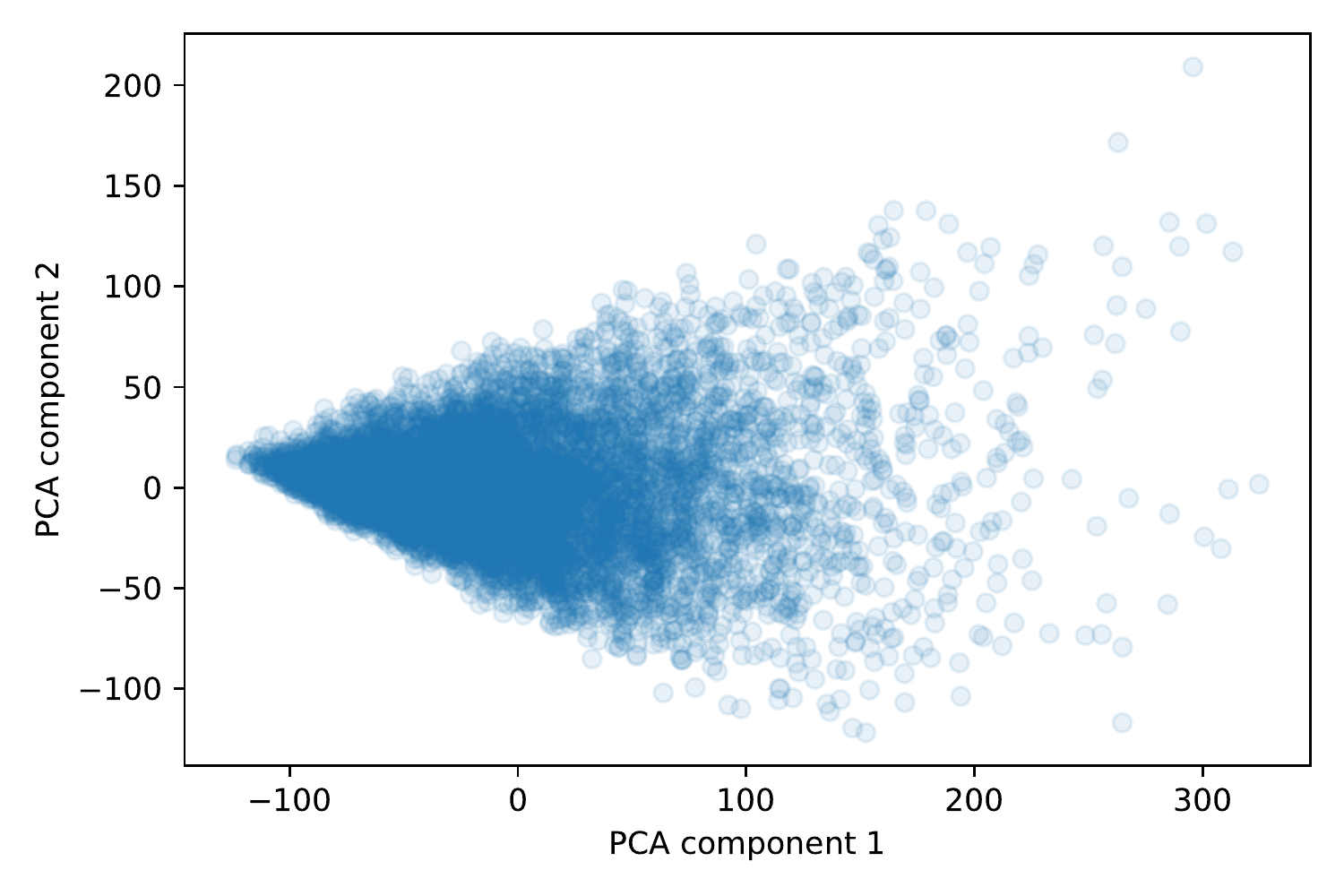}
    	\caption{Coprime 5-vectors}\label{CoprimePCA}
    \end{subfigure} \\ 
	\begin{subfigure}{0.45\textwidth}
		\centering
		\includegraphics[width=\textwidth]{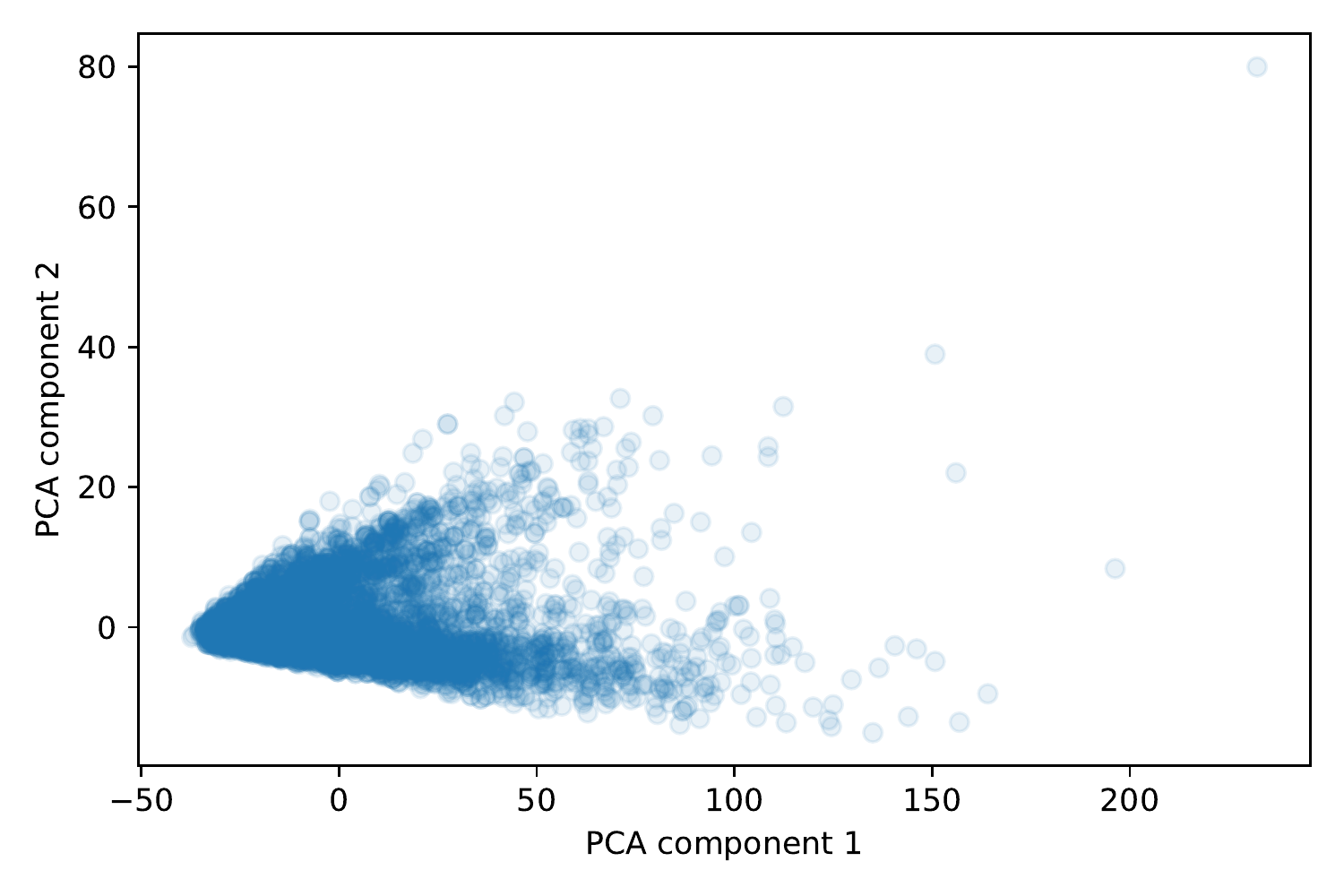}
		\caption{Transverse 5-vectors}\label{TransversePCA}
	\end{subfigure} 
	\begin{subfigure}{0.45\textwidth}
		\centering
		\includegraphics[width=\textwidth]{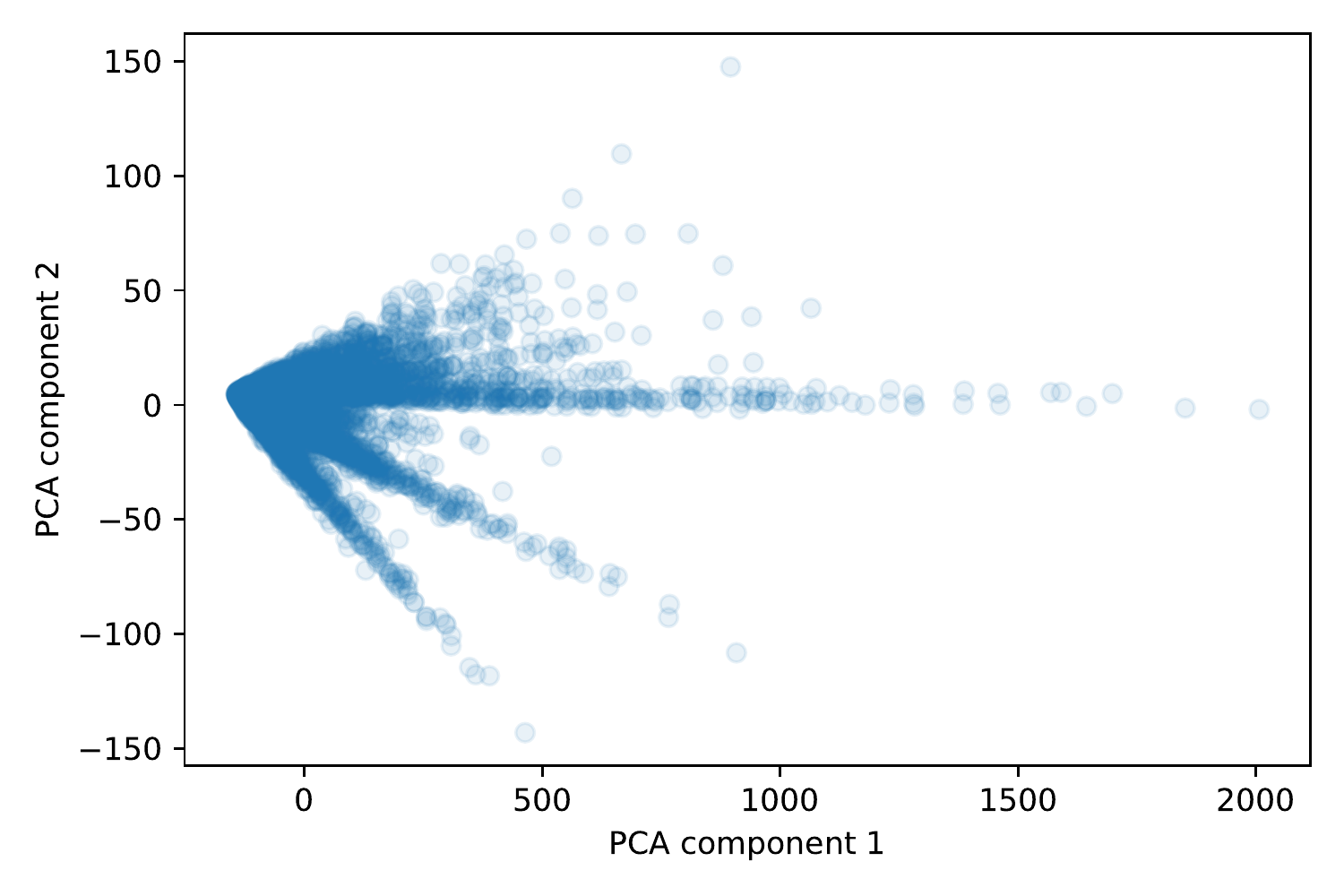}
		\caption{CY 5-vectors}\label{CYPCA}
	\end{subfigure} 
\caption{2-dimensional PCA plots for the 4 considered datasets. As more of the conditions are added, more structure appears, in particular there is some form of distinct class separation for the CY weights.}\label{PCA_CYDatasets}
\end{figure}

For the 4 datasets considered: (a) random 5-vectors, (b) coprime 5-vectors, (c) transverse 5-vectors, (d) CY dataset; the most dominant principal component in each case had normalised eigenvalue entries of: 0.76, 0.75, 0.91, 0.98 respectively.
These indicate that a 1-dimensional projection would be sufficient for preserving the majority of the data, however to consistently account for $>0.9$ of the data information a 2-dimensional projection was used in each case.
Despite the latter entries dominating these principal components as expected, the lack of a block diagonal form for the covariance matrices indicated a slightly more subtle structure using all the vector information.
These projections are shown in Figure \ref{PCA_CYDatasets}, where the cone-like structure is an artefact of the vector sorting, however the separation into linear clusters in the CY dataset case is unexplained and surprising.

\begin{figure}[tb]
    \centering
    \includegraphics[width=0.5\textwidth]{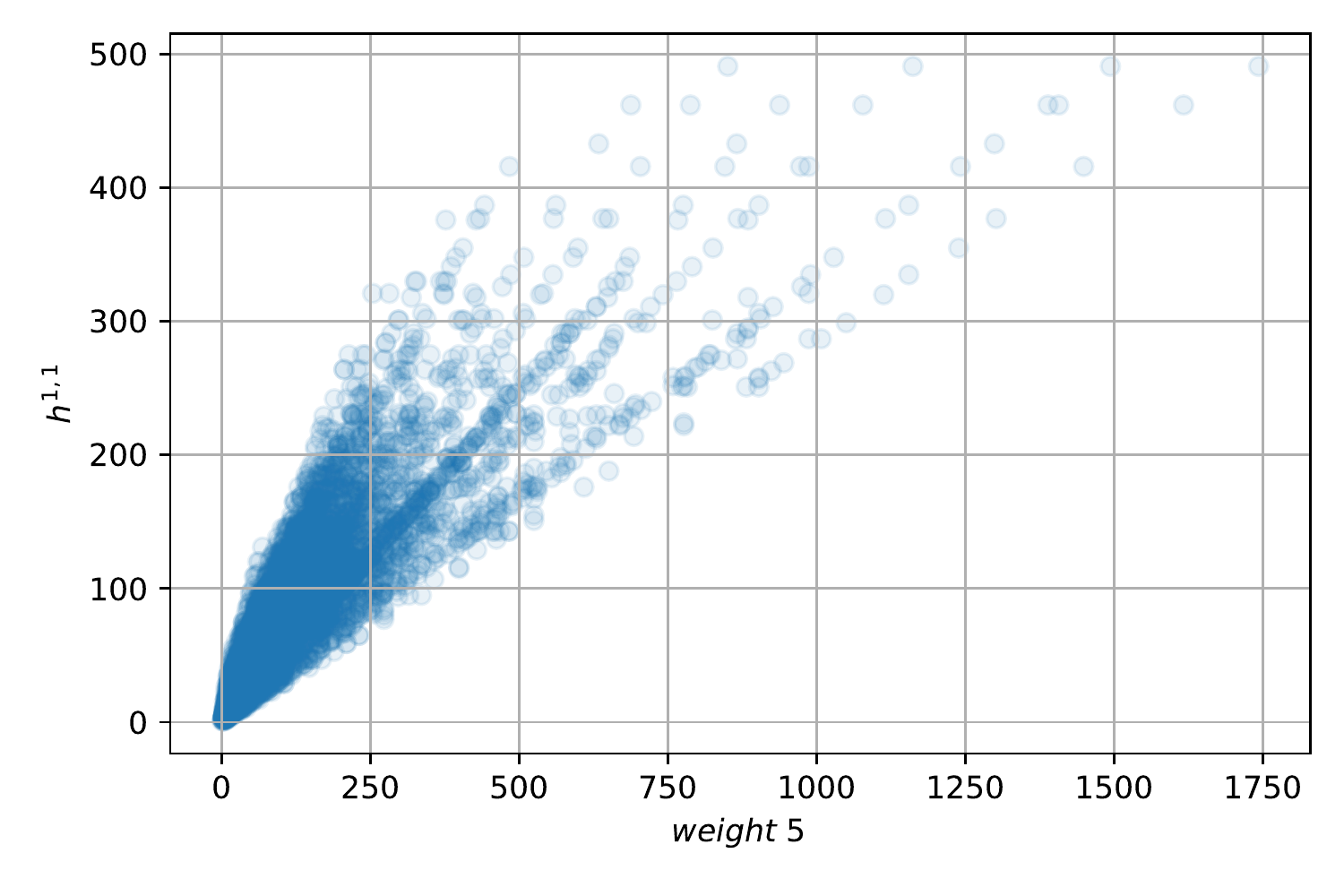}
    \caption{The equivalent linear clustering behaviour of the Calabi-Yau weights with the $h^{1,1}$ Hodge number corroborated by the PCA results.}
    \label{CY_weighthodgeplot}
\end{figure}

This linear clustering of the CY weights was a novel fundamental structure, and its lack of occurrence in the conjugate non-CY datasets highlights its uniqueness to the CY property.
Further analysis in this work showed consistent linear correlation to the cluster separation formed by plotting the weights (especially the largest weight), against the hypersurface's $h^{1,1}$ value, as shown in Figure \ref{CY_weighthodgeplot}.
These PCA results hence nicely corroborate the existence of a more direct connection to the hypersurface's $h^{1,1}$ value within these linear clusters, and petitions for investigation into the fundamental structure behind this apparently simple linear clustering behaviour.

\subsection{t-SNE}\label{sec:tSNE}

t-distributed stochastic neighbour embedding, or t-SNE \cite{Maaten2008}, is a dimensionalilty reduction method that reduces high-dimensional data by embedding it in a low-dimensional space. It calculates similarities between data points in the high dimensional space and in the low dimensional space and tries to minimise the divergence between the two. The process can be broken down into three steps:
\begin{enumerate}
    \item Given a set of $N$ points, $\textbf{x}_{1},...,\textbf{x}_{N}$ in high dimensional space, compute the similarities between pairs of points. The similarity of data point $\textbf{x}_{i}$ to data point $\textbf{x}_{j}$ is defined as:
    \begin{equation}
        p_{ij} = \frac{p_{j|i}+p_{i|j}}{2N}\,,
    \end{equation}
    where $p_{j|i}$ is the conditional probability that $\textbf{x}_{i}$ would pick $\textbf{x}_{j}$ as its neighbour. This probability is proportional to the probability density under a Gaussian centred at $\textbf{x}_{i}$:
    \begin{align}
    p_{j|i} = 
        \begin{cases}
            \frac{\text{exp}(-||\textbf{x}_{i} - \textbf{x}_{j}||^{2} /2\sigma_{i}^{2})}{\sum_{k \neq i}{\text{exp}(-||\textbf{x}_{i} - \textbf{x}_{k}||^{2} /2\sigma_{i}^{2})}} & \text{if } i \neq j \\
            0 & \text{if } i = j
        \end{cases}\;,
    \end{align}
    where $\sigma_i$ are the standard deviations. 
    \item In the low dimensional space compute the similarity measures between data points, but instead of using a Gaussian distribution use a Student's t-distribution with one degree of freedom. In this case, the similarity of data point $\textbf{y}_{i}$ to data point $\textbf{y}_{j}$ is defined as:
    \begin{align}
    q_{ij} = 
        \begin{cases}
            \frac{(1+||\textbf{y}_{i} - \textbf{y}_{j}||^{2})^{-1}}{\sum_{k \neq l}{(1+||\textbf{y}_{k} - \textbf{y}_{l}||^{2} )^{-1}}} & \text{if } i \neq j \\
            0 & \text{if } i = j
        \end{cases}\;.
    \end{align}
    \item Compute the Kullback-Liebler divergence between the two probability distributions $p_{ij}$ and $q_{ij}$:
    \begin{align}
        D_{KL}(P||Q) = \sum_{i \neq j}{p_{ij}\log{\frac{p_{ij}}{q_{ij}}}}\;.
    \end{align}
    This divergence is minimised with respect to the data points $\textbf{y}_{i}$ using gradient descent. 
\end{enumerate}

\begin{figure}[tb]
    \centering
    \includegraphics[width=0.5\textwidth]{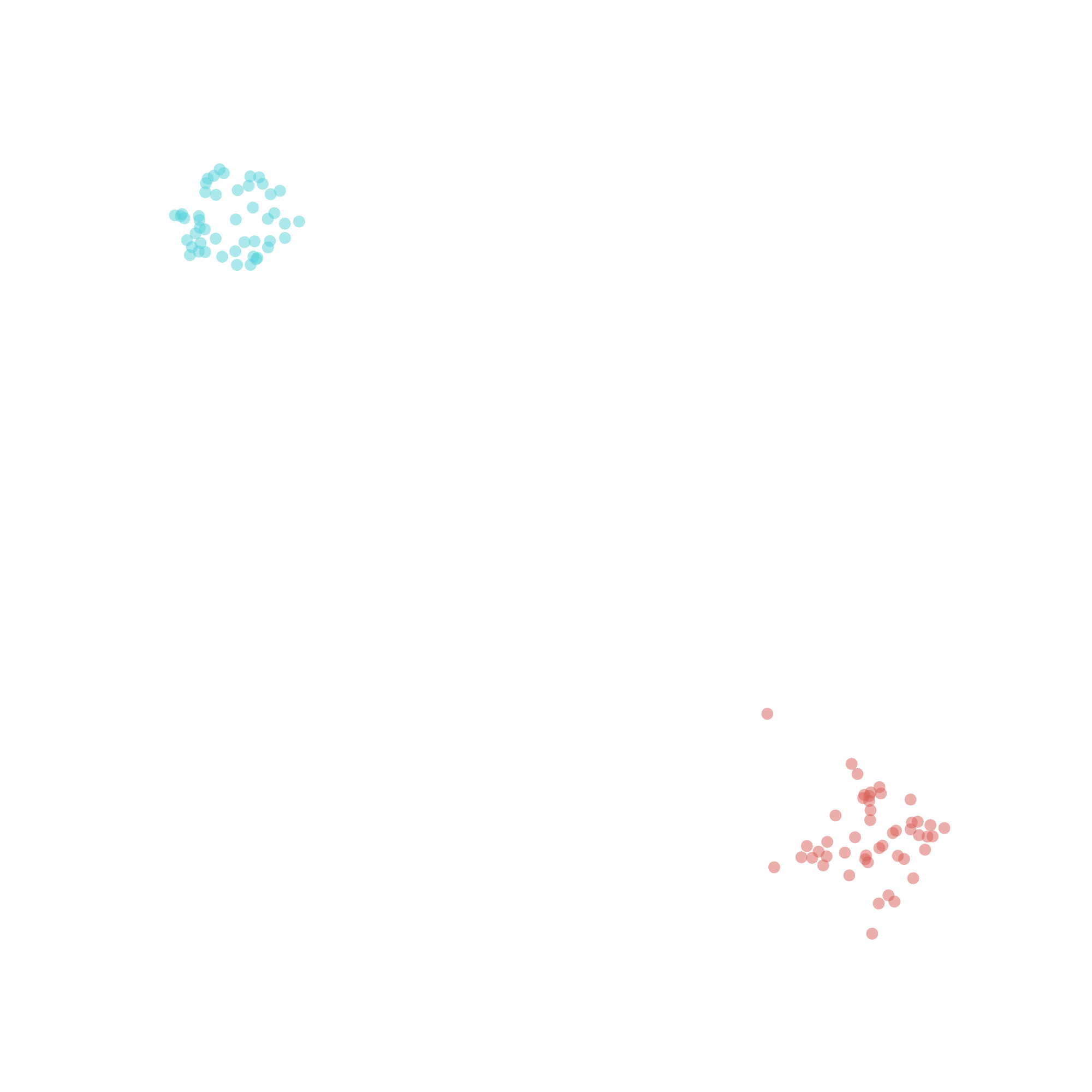}
    \caption{t-SNE plot of the 10-dimensional web embeddings generated by the SNN, reduced to 2 dimensions.}
    \label{fig:tSNE_results}
\end{figure}

\paragraph{Example: Brane Webs}\mbox{}\\ 
In §\ref{sec:snn} we discussed how an SNN was used in \cite{Arias-Tamargo:2022qgb} to predict whether two brane webs are equivalent. Recall, that two webs are equivalent if they share the same self-intersection number \eqref{eq:self_intersection}, total monodromy \eqref{eq:total_monodromy} and asymptotic charge invariant \eqref{eq:asymptotic_charge}. It would nice to visualise the embeddings of the webs in the test set to see whether equivalent webs appear to cluster together as we would like them to. The dimension of the embedding space used in the SNN was $d=10$ and so t-SNE is applied to the resulting embeddings in order to reduce the embeddings down to 2 dimensions. The result are shown in Figure \ref{fig:tSNE_results}. We can see clearly that the embeddings coming from the two equivalence classes $\textbf{X}_{1}\cup\textbf{X}_{2}$ cluster together into two distinct sets. This aligns with the pairwise equivalence prediction result we saw earlier in §\ref{sec:snn}.

\subsection{K-Means Clustering}\label{sec:kmeans} 

$k$-means clustering is a clustering algorithm that groups data points into $k$ clusters, where each data point belongs to the cluster with the nearest centroid. The objective is to minimise the sum of squared distances between the data points and the nearest centroid, known as the inertia $\mathcal{I}$. In other words, given a set of data points $(\textbf{x}_{1},...,\textbf{x}_{n})$, the $k$-means clustering algorithm partitions the data points into $k$ sets $\textbf{S}=(S_{1},...,S_{k})$ so as to minimise the sum:
\begin{align}
    \mathcal{I} = \sum_{i=1}^{k}{\sum_{\textbf{x} \in S_{i}}{||\textbf{x} - \mu_{i}||^2}}\,,
\end{align}
where $\mu_{i}$ denotes the centroid of the set $S_{i}$.

The first step is to choose $k$, i.e. the number of clusters. The next step is to randomly select $k$ points from the data and label them as centroids. One then assigns each data point to the cluster with the closest centroid. Once data points are sorted into clusters, the new centroid of each cluster is computed and again the data points are assigned to the nearest cluster. This two step process of computing the new centroid and then assigning data points to clusters is repeated until the centroids do not change their position.
Where this input $k$ number of clusters is unknown, an elbow method can be implemented whereby the clustering is run until completion for a range of integer values and the final inertia is added to a penalty factor for the number of clusters used; leading to a convex optimisation to easily extract the discrete minimum $k$ value for the optimum number of clusters.

One common metric used to evaluate clustering performance is Rand index (RI). RI measures the similarity between two partitions by considering all pairs of elements and counting pairs where both elements are assigned to the same subset, or different subsets. The two partitions compared are the predicted partition and the true partition. Let $X=\{x_1,...,x_n\}$ be a set of $n$ elements and $A=\{A_1,...,A_r\}$, $B=\{B_1,...,B_s\}$ be two partitions of $X$ into $r$ and $s$ subsets respectively. Then the Rand index is given by 
\begin{align}\label{eq:RI}
    RI = \frac{a+b}{a+b+c+d}\;,
\end{align}
where 
\begin{itemize}
    \item $a$ is the number of pairs of elements in $X$ that are in the same subset in $A$ and the same subset in $B$.
    \item $b$ is the number of pairs of elements in $X$ that are in different subsets in $A$ and different subsets in $B$.
    \item $c$ is the number of pairs of elements in $X$ that are in the same subset in $A$ and different subsets in $B$.
    \item $d$ is the number of pairs of elements in $X$ that are in different subsets in $A$ and the same subset in $B$.
\end{itemize}
Therefore RI exists in the range [0,1], where 1 corresponds to a perfect agreement between the partitions and 0 corresponds to complete disagreement.

\paragraph{Example: Brane Webs}\mbox{}\\
Since the idea of an SNN is to map elements in a dataset to an embedding space, such that similar elements are mapped close together in the embedding space and dissimilar elements are mapped far apart, it makes sense to apply $k$-means cluster on the embeddings in order to group the data into equivalence classes. This is precisely what the authors in \cite{Arias-Tamargo:2022qgb} did to the embeddings of 5-brane webs produced by the SNN, see §\ref{sec:snn} for more details. Setting $k=2$ the Rand index score of the results of $k$-means clustering was $1.0000\pm0.0000$ which means that the $k$-means clustering matched perfectly with the true clustering. These results also support the conclusions made from the t-SNE plot in Figure \ref{fig:tSNE_results}.

\paragraph{Example: Calabi-Yau Threefolds}\mbox{}\\ 
The linear clustering behaviour observed in the PCA and Hodge plots, Figures \ref{PCA_CYDatasets} and \ref{CY_weighthodgeplot} in §\ref{sec:pca}, advocates for the application of unsupervised clustering methods.
The authors of \cite{Berman:2021mcw} chose to use $k$-means to analyse this clustering behaviour.  

Since each cluster in Figure \ref{CY_weighthodgeplot} appears to lie on a line through the origin, the data can be projected down onto the ratio $h^{1,1}/w_5$ to create a 1-dimensional clustering problem for the K-Means algorithm. Running the $k$-means elbow method dictates that there is an optimum number of 10 clusters, matching the observed number from the plots.
In addition, the normalised inertia value with respect to the number of vectors considered and the range of ratios across this dataset took a value of $0.00084$; indicating that each ratio was $<0.1\%$ of the ratio-range away from its nearest cluster, corroborating the existence of a linear clustering in these CY 5-vectors.

\subsection{Topological Data Analysis}\label{sec:tda} 

Data visualisation is useful for the development of one's intuition when working with a dataset. However, our ability to picture data is limited to at most three dimensions. This motivates the use of techniques that project data to a lower dimension, such as PCA (§\ref{sec:pca}) and t-SNE (§\ref{sec:tSNE}). Any data projection is inevitably going to lose some information and hence some perspective for data interpretation. 

Topological data analysis (TDA) is the name given to the field of techniques that analyse the topological properties of a higher-dimensional datasets without projection. This direct analysis is in a sense more physical, and naturally evades the errors that occur with more standard visitation. 
The most prominent technique within TDA uses persistent homology. Here, the $n$-dimensional data in $\mathbb{R}^n$ space is used to build a filtration of simplicial complexes on which persistent homology can be applied. The computation process initially generates an equivalent simplicial complex analogous to the data distribution with a 0-simplex (point) for each datapoint. Then it centres an $n$-dimensional ball on each data point of radius $\delta$ and smoothly increases the value of $\delta$ from 0 to $\infty$. At each new $\delta$ where there is a new intersection of $k \leq n$ balls an equivalent $k$-simplex is drawn in the simplicial complex between the respective 0-simplices; creating a discrete filtration of complexes over the range of $\delta$ values.

The algorithm keeps track of the persistent features of each dimension throughout the filtration, whereby a $H_k$ feature is a set of $k$-simplices which do not form the boundary of a set of $(k+1)$-simplices. The dominant persistent features which exist for a large range of $\delta$ values indicate intrinsic structure to the data. These features can be represented on a persistence diagram, which is a 2-dimensional plot with coordinates (birth, death), where `birth' is the $\delta$ value the feature was created in the filtration, and `death' the $\delta$ value that it disappeared / was filled.

\paragraph{Example: Amoebae Images}\mbox{}\\ 
Considering amoebae in dimension $n$, the most relevant topological properties are the existence of $H_{n-1}$ features counted by the relevant Betti number $b_{n-1}$.
In the 2-dimensional examples considered in \cite{Bao:2021olg} these are $H_1$ features which are 2-dimensional holes, for the 3-dimensional examples considered in \cite{Chen:2022jwd} these are $H_2$ features which are 3-dimensional cavities.
Their existence is identified through TDA's persistent homology by the presence of persistent features far from the diagonal of the amoeba's persistence diagram.

\begin{figure}[!tb]
    \centering
    \begin{subfigure}{0.45\textwidth}
        \centering
        \includegraphics[width=0.75\textwidth]{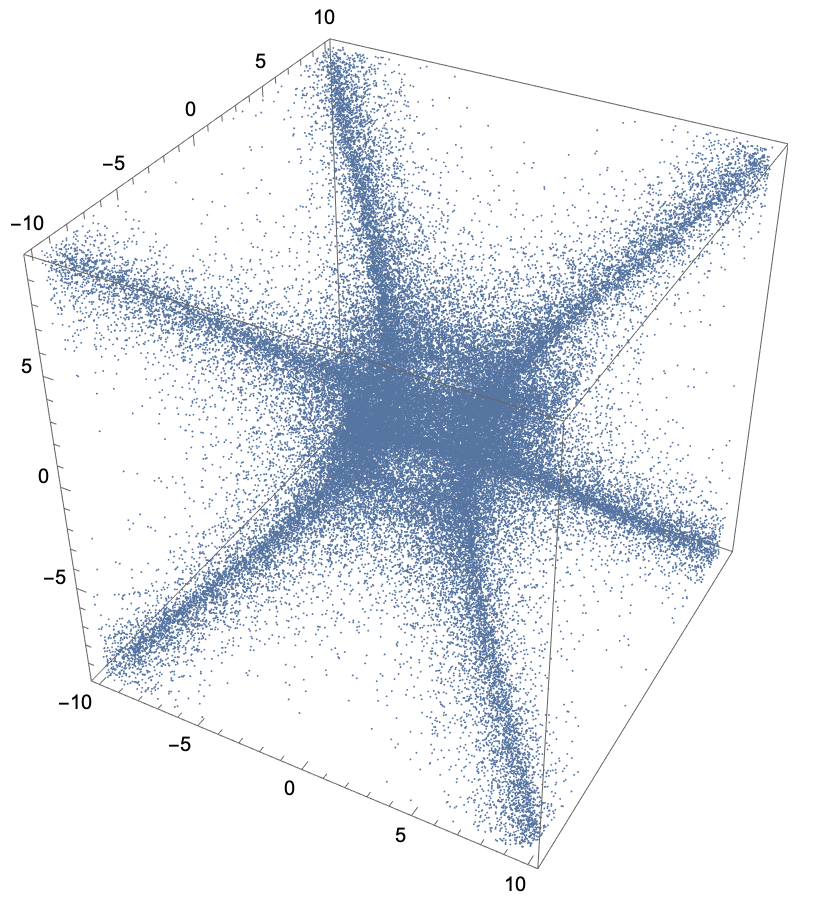}
        \caption{Monte Carlo sampled amoeba}\label{amb_3dexample}
    \end{subfigure}
    \begin{subfigure}{0.45\textwidth}
        \centering
        \includegraphics[width=0.9\textwidth]{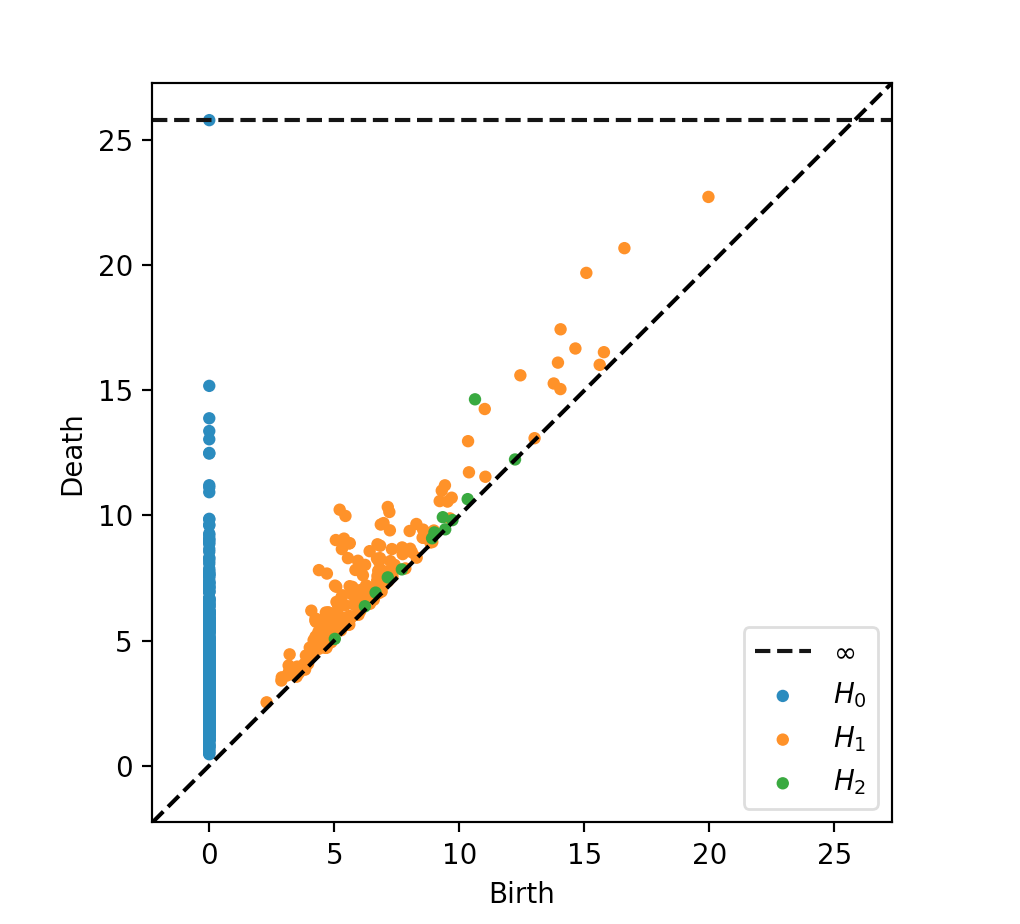}
        \caption{Amoeba persistence diagram}\label{amb_persistence}
    \end{subfigure}
    \caption{For the $\mathbb{P}^1 \times \mathbb{P}^1 \times \mathbb{P}^1$ geometry, the respective 3-dimensional amoeba is shown in (a) where the points are sampled from the complex polynomial and projected to the amoeba. In (b) the equivalent persistence diagram is shown for the same sampling, where the green dot far from the diagonal ($\sim (10.6,14.6)$) indicates a $H_2$ feature.}\label{amb_tda_example}
\end{figure}

In \cite{Bao:2021olg} persistent homology was used to reanalyse a set of misclassified amoeba, and the procedure managed to classify the genus more accurately than the CNN architectures on these more difficult examples.
In \cite{Chen:2022jwd} the persistent homology process was tested more thoroughly as the automated computation measure for determining the label ahead of the CNN learning.
Although the computation of the persistent features is expensive, the procedure is simply automatable, and thus more practical for implementation than the heuristic identification of the suitable bounds using lopsidedness.
An example Monte Carlo sampled 3-dimensional amoeba image is given in Figure \ref{amb_tda_example}, along with its respective persistence diagram -- demonstrating the correct identification of a $H_2$ feature where $b_2=1$.

\section{Conclusion}\label{sec:conc} 
Through a selection of research works, this chapter has exemplified the successes of machine learning methods on problems within physics and geometry.As computational power and proficiency continues to develop, the abundance of mathematical data will further the demands for statistical methods of analysis, particularly the range of techniques available under the umbrella of machine learning.

The techniques of focus here have been those of simpler structure, and motivate the use of more advanced (and appropriately specialised) methods for development of the analysis, and extraction of concrete mathematical insights. 
Whilst this work highlights the initial successes of ML in the field of physics, it is also worth noting that there are now multiple programs of work to apply ideas from quantum field theory and high-energy physics to understand the ubiquitous successes of NNs \cite{Halverson:2020trp,Berman:2022mak,Berman:2022uov}.

As mathematical physics and machine learning continue to intertwine, their breadth of cross-application will no doubt lead to exciting insightful mutual development.

\section*{Acknowledgement}
YHH would like to thank STFC for grant ST/J00037X/2.
E.~Heyes would like to thank SMCSE at City, University of London for the PhD studentship, as well as the Jersey Government for a postgraduate grant.
E.~Hirst would like to thank STFC for a PhD studentship.

\addcontentsline{toc}{section}{References}
\printbibliography
\end{document}